\documentclass[aps,pra,twocolumn,superscriptaddress]{revtex4-1}

\usepackage{mathtools}
\usepackage[usenames,dvipsnames,table]{xcolor}
\usepackage{float}
\usepackage{amssymb,bm}
\usepackage{tabularx, booktabs}
\usepackage{tikz,pgfplots}
\usepackage{amsmath,amsfonts,mathrsfs,amssymb}
\usepackage{url}
\usepackage{graphicx}
\usepackage{epsfig}
\usepackage{color}
\usepackage{bm}
\usepackage{array}
\usepackage[colorlinks=true, linkcolor=blue, citecolor=dred]{hyperref}
\usepackage{csquotes}
\usepackage{xpatch}
\usepackage{xspace}
\usepackage[normalem]{ulem} 
\usepackage{physics}
\usepackage{appendix}
\newcolumntype{Y}{>{\centering\arraybackslash}X}
\newcommand{\etal}{{\it{et al.}}}
\newcommand{\ie}{{\it{i.e.}}}

\newcommand{\bfr}{\mathbf{r}}

\newcommand{\bfn}{\mathbf{n}}

\newcommand{\bepsi}{\bm{\varepsilon}}
\newcommand{\psh}[2]{\ensuremath{\langle #1|#2\rangle}\xspace}
\newcommand{\manu}[1]{{\textcolor{dgreen}{ Manu: #1 }} }

\definecolor{dgreen}{rgb}{0,.5,0}
\definecolor{dblue}{rgb}{0,0,.5}
\definecolor{dred}{rgb}{0.5,0,.5}
\newcommand{\be}{\begin{eqnarray}}
\newcommand{\ee}{\end{eqnarray}}
\newcommand{\exact}{{``exact'' }}

\begin{document}

\title{
Reduced density matrix functional theory from an {\it ab
initio} seniority-zero wave function: Exact and approximate formulations
along adiabatic connection paths
%
}

\author{Bruno Senjean}
\thanks{Corresponding author}
\email{bruno.senjean@umontpellier.fr}
\affiliation{ICGM, Universit\'{e} de Montpellier, CNRS, ENSCM, Montpellier, France}

\author{Saad Yalouz}
\affiliation{Laboratoire de Chimie Quantique,
Institut de Chimie, CNRS/Universit\'{e} de Strasbourg,
4 rue Blaise Pascal, 67000 Strasbourg, France}

\author{Naoki Nakatani}
\affiliation{Department of Chemistry,
Graduate School of Science and Engineering,
Tokyo Metropolitan University,
1-1 Minami-Osawa, Hachioji,
Tokyo 192-0397, Japan}
\affiliation{Institute for Catalysis,
Hokkaido University,
N21W10 Kita-ku, Sapporo,
Hokkaido 001-0021, Japan}

\author{Emmanuel Fromager}
\affiliation{Laboratoire de Chimie Quantique,
Institut de Chimie, CNRS/Universit\'{e} de Strasbourg,
4 rue Blaise Pascal, 67000 Strasbourg, France}

\date{\today}

\begin{abstract}

Currently, there is a growing interest in the development of a new hierarchy of methods 
based on the concept of seniority, which has been introduced quite
recently in quantum chemistry.
Despite the enormous potential of these methods, the accurate description of both dynamical and static correlation effects
within a single and in-principle-exact approach remains a challenge.
In this work, we propose an alternative formulation of reduced density-matrix functional theory (RDMFT)
where the (one-electron reduced) density matrix is mapped onto an {\it
ab initio} seniority-zero wave function. In this theory, the exact 
natural orbitals and their occupancies are determined self-consistently
from an effective seniority-zero calculation. The latter involves
a universal higher-seniority density matrix functional for
which an adiabatic connection (AC) formula is derived and implemented under specific
constraints that are related to the density matrix.
The pronounced curvature of the (constrained) AC integrand, which is numerically
observed in prototypical hydrogen chains and the Helium
dimer, indicates that a description of higher-seniority correlations
within 
second-order perturbation theory is inadequate in this context. Applying multiple linear
interpolations along the AC or connecting second-order perturbation theory to a
full-seniority treatment {\it via} Pad\'{e} approximants are better
strategies. Such
information is expected to serve as a guide in the future design of higher-seniority
density-matrix functional approximations.



\end{abstract}

\maketitle

\section{Introduction}

The accurate description of chemical compounds
which manifest both weak and strong correlation effects
(also referred to as dynamical and static correlation effects,
respectively~\cite{becke2013density,stein2016delicate})
remains
a long-standing challenge in the quantum chemistry 
community.
One angle of attack to tackle the
electronic structure problem is to
construct orbital-based wave functions, expressed
as linear combination of Slater determinants.
Most quantum chemistry methods
are built upon Hartree--Fock (HF) theory and include corrections
from excited-state configurations, thus leading to
configuration interaction (CI)
and coupled cluster (CC) methods~\cite{helgaker2014molecular}. 
If all configurations
are considered, it is called the full configuration interaction (FCI)
and
all the correlation effects are captured. Unfortunately,
the computational
cost associated to FCI scales exponentially with the system size,
such that the configuration space is commonly truncated
to single and double excited-state configurations (CISD and CCSD).
Despite the success of those approximations (CC is
considered as the ``gold standard'' method in quantum 
chemistry~\cite{bartlett2007coupled,lyakh2012multireference}),
strong correlation effects are still insufficiently described.
A similar conclusion holds for the 
Kohn--Sham density functional theory (KS-DFT)~\cite{hktheo,KS}
whose density functional approximations are not able to describe
strong correlation effects~\cite{cohen2011challenges,
swart2013spin,swart2016spinning}.
Instead of using the electronic density as a basic variable, one
can consider the one-particle reduced density matrix (1RDM),
thus leading to the reduced density matrix functional theory 
(RDMFT)~\cite{gilbert1975hohenberg,donnelly1978elementary,
donnelly1979fundamental,pernal2015reduced}. Several
1RDM functionals based on the occupation numbers of the 
natural orbitals
have been developed~\cite{muller1984explicit,
goedecker1998natural,holas1999properties,
buijse2002approximate,gritsenko2005improved,
sharma2008reduced,lathiotakis2009density,lathiotakis2009functional,
marques2008empirical,rodriguez2017comprehensive},
in particular the Piris natural orbital functionals
PNOF$i$ ($i=1,7$)~\cite{piris2003one,leiva2005natural, 
piris2006new,piris2007dispersion,piris2009iterative,piris2011natural,
piris2013natural,piris2014interacting,
piris2014perspective,ramos2015h4,lopez2015performance,
mitxelena2017performance}.
In order to treat both the static and dynamical correlations
simultaneously, several hybrid
schemes, where wave function theory is combined with
the aforementioned reduced quantity theories~\cite{savinbook,toulouse2004long,fromager2015exact,
senjean2019projected,libisch2014embedded,
li2014multiconfiguration,gagliardi2016multiconfiguration,
mostafanejad2018combining,mostafanejad2020global,
lee2019projection,mosquera2019domain},
have been proposed. 
Quite recently, Pernal~\cite{pernal2018electron} 
also suggested to use the adiabatic 
connection (AC) formalism to recover the electron correlation that is
missing in multiconfigurational wave functions~\cite{pernal2018exact,pastorczak2019capturing,
maradzike2020reduced,drwal2022efficient}.

Post-HF methods are adequate for treating weakly correlated
systems. When it comes to describing strongly correlated systems, a more
appealing strategy consists in dropping the orbital picture and,
instead, constructing wave functions as products of geminals (\ie, pairs of electrons)~\cite{bardeen1957superconductivity,hurley1953molecular,
kutzelnigg1964direct,coleman1965structure,silver1969natural,
surjan1999introduction,rassolov2002geminal,surjan2016geminal,
fecteau2021richardson}.
A hierarchy of these methods can be found in
Ref.~\cite{johnson2013size}.
Starting from the idea of pairing electrons, one can divide the entire
many-electron Hilbert space into subspaces
by using the 
concept of seniority number. The latter corresponds to the number of unpaired
electrons in a determinant~\cite{bytautas2011seniority,bytautas2018seniority}.
If all electrons are paired (\ie, all spatial orbitals are either doubly
occupied or unoccupied), the 
wave function is referred to as {\it seniority-zero} wave function. It
is described exactly by performing a
doubly occupied configuration interaction
(DOCI) calculation~\cite{weinhold1967reduced} where the seniority-zero
part of the 
Hamiltonian~\cite{limacher2016new} is diagonalized.\\

Even though seniority-zero wave functions can describe accurately strong
(static) correlation effects, their evaluation is computationally
demanding and a proper description of dynamical correlation effects requires taking into account
higher seniorities~\cite{gunst2021seniority} or mixing excitation
degrees and seniority numbers~\cite{kossoski2022hierarchy}.
Other promising approaches that retain the
computational cost of a mean-field and are adequate for the treatment of
strong correlation have been proposed,
like the so-called
antisymmetric product of one-reference-orbital geminals
(AP1roG)~\cite{johnson2013size,limacher2013new,tecmer2014assessing,
boguslawski2014efficient}
equivalent to the pair-coupled cluster theory 
(pCCD)~\cite{stein2014seniority,henderson2015pair},
or more recently the pair parametric two-electron
reduced density matrix approach~\cite{vu2020size}.
Open-shell extensions of pCCD have also been developed
in Ref.~\onlinecite{boguslawski2021open}.
Orbital optimization is essential in these approaches for the accurate description of both ground~\cite{limacher2014influence,boguslawski2014efficient,
boguslawski2014nonvariational,
boguslawski2014projected,stein2014seniority} and excited states~\cite{kossoski2021excited,marie2021variational,
boguslawski2016targeting,boguslawski2018targeting}.
Despite promising results,
AP1roG and related geminal-based wave functions
are not adequate to capture dynamical
correlation~\cite{boguslawski2016analysis}.
Indeed, dynamical correlation is not described by 
electron-pair states but by higher-seniority contributions. The latter
can be determined from perturbation 
theory~\cite{roeggen1999extended,
rosta2000interaction,
rosta2002two,rassolov2004geminal,
kobayashi2010generalized,tarumi2012generalized,
piris2013interpair,
limacher2015orbital,
boguslawski2017benchmark,piris2017global,brzek2019benchmarking},
coupled-cluster theory~\cite{brzek2019benchmarking,kutzelnigg2012separation,
zoboki2013linearized,hollett2020capturing,margocsy2020calculation,
nowak2021orbital,henderson2014seniority,boguslawski2015linearized,
leszczyk2021assessing},
the extended random phase approximation~\cite{pernal2014intergeminal,
margocsy2018multiple},
DFT~\cite{garza2015synergy,garza2015range,hollett2016cumulant}, or the AC formalism~\cite{vu2019adiabatic}.
In the spirit of post-HF methods,
post-Antisymmetrized Geminal Power approaches are also
promising, not only on classical computers~\cite{henderson2019geminal,henderson2020correlating,
dutta2020geminal,khamoshi2021exploring}, but also on quantum
computers~\cite{khamoshi2020correlating}.\\

In this work, we propose to formally exactify seniority-zero wave function
calculations in the context of {\it reduced density matrix functional
theory} (RDMFT). More precisely, by analogy with KS-DFT, we will map the
1RDM onto a reference {\it ab initio} seniority-zero wave function. In
this context, the
higher-seniority correlation effects will be described by a density
matrix functional. The paper is organized as follows. In Sec.~\ref{sec:theory}, we present an exact reformulation of RDMFT as an 
effective seniority-zero theory. The self-consistent calculation of both the
natural orbitals and their occupancies will be discussed in this
context. In order to give further insight into
the complementary higher-seniority density-matrix correlation
functional, which is a central object in the theory, two different adiabatic connection
paths are constructed in Sec.~\ref{sec:AC_formalism}. Details about the
computational implementation of the adiabatic connections are provided
in Sec.~\ref{sec:comput_details}, which is followed by a detailed
discussion in Sec.~\ref{sec:discussion} of the results obtained for the prototypical H$_4$ and H$_8$
chains, as well as the Helium dimer. Conclusions and perspectives are
given in Sec.~\ref{sec:conclusions}.

\section{Theory}\label{sec:theory}

\subsection{Seniority-zero wave function and
Hamiltonian}\label{subsec:seniority_zero}

The concept of seniority has been introduced recently in quantum chemistry
by Scuseria and co-workers~\cite{bytautas2011seniority}.
It consists in partitioning the FCI
Fock space into subspaces that are defined from the following seniority number operator:
\begin{eqnarray}\label{seniority_op}
\hat{\Omega} = \sum_{p} \left( \hat{n}_{p\uparrow} + \hat{n}_{p\downarrow} - 2 \hat{n}_{p\uparrow} \hat{n}_{p\downarrow}\right),
\end{eqnarray}
where $\hat{n}_{p\sigma}\equiv \hat{c}_{p\sigma}^\dagger
\hat{c}_{p\sigma}$ [$\sigma=\uparrow,\downarrow$] is a
spin-orbital occupation operator written in second quantization. As readily seen from
Eq.~(\ref{seniority_op}), the seniority operator $\hat{\Omega}$ simply counts the number of unpaired electrons in a given
configuration.
It has been shown that the
seniority-zero sector, which is denoted $\mathcal{S}_0$ in the
following and for which $\langle \hat{\Omega} \rangle_{\mathcal{S}_0}
= 0$, contributes significantly to the description
of static correlation~\cite{bytautas2011seniority}.
In the DOCI
approximation~\cite{weinhold1967reduced}, FCI is applied to the
seniority-zero subspace, which means that determinants with
empty or doubly occupied orbitals only are considered in the calculation. In such a case,
the wave function ansatz reads 
\be\label{eq:S0_parameterization}
\mathcal{S}_0\ni\Psi
\equiv \sum_{p_1<\ldots< p_{N/2}} C_{p_1\ldots p_{N/2}}
\prod^{N/2}_{i=1}\hat{c}_{p_i\uparrow}^\dagger\hat{c}_{p_i\downarrow}^\dagger\ket{\rm vac}
,
\ee
where $\ket{\rm vac}$ denotes the vacuum state and $N$ is the (even) number of
electrons. Note that the 1RDM of a seniority-zero wave function is
diagonal in its orbital basis: 
\be\label{eq:diag_S0_RDM}
{D}_{pq}^{\Psi}=\sum_\sigma
\bra{\Psi}\hat{c}_{p\sigma}^\dagger\hat{c}_{q\sigma}\ket{\Psi}\overset{\Psi\in\mathcal{S}_0}{=}
 \delta_{pq}n^\Psi_p,
\ee
where
\be
n^\Psi_p=2\sum_{p_1<\ldots< p_{N/2}} \abs{C_{p_1\ldots
p_{N/2}}}^2\sum^{N/2}_{i=1}\delta_{pp_i}.
\ee
In other words, these orbitals are the seniority-zero
natural orbitals, by construction. By analogy with the {\it multiconfigurational self-consistent
field} (MCSCF) method~\cite{helgaker2014molecular}, they can be optimized
variationally, in addition to the DOCI coefficients
$\left\{C_{p_1\ldots p_{N/2}}\right\}$~\cite{vu2019adiabatic}. In the
present work, we aim at deriving an in-principle-exact variational
energy expression such that the minimizing seniority-zero orbitals match the exact natural orbitals of the true
(higher-seniority) physical many-body wave function. As briefly sketched
in the following and further formalized in the next subsections, 
we need for that purpose to introduce a density matrix functional
correction to the (approximate) seniority-zero energy.\\

Let us start with the general ({\it ab initio}) second-quantized
Hamiltonian expression,
\be
\hat{H}&=&\sum_{pq}\sum_\sigma h_{pq}\hat{c}_{p\sigma}^\dagger\hat{c}_{q\sigma}
+\hat{W}_{\rm ee},
\\
\label{eq:full_2e_repulsion_unrotated_basis}
\hat{W}_{\rm ee}&=&\dfrac{1}{2}\sum_{pqrs}\psh{pq}{
rs}                                                                                        \sum_{\sigma\sigma'}\hat{c}^\dagger_{p\sigma}\hat{c}^\dagger_{q\sigma'}\hat{c}_{s\sigma'}\hat{c}_{r\sigma},
\ee
where $h_{pq}=\bra{\varphi_p}\hat{h}\ket{\varphi_q}=\int d\bfr\, \varphi_p(\bfr)\left[-\frac{1}{2}\nabla_{\bfr}^2+v_{\rm
ne}(\bfr)\right]\varphi_q(\bfr)$ and $\psh{pq}{
rs}=\int\int
d\bfr
d\bfr'\varphi_p(\bfr)\varphi_q(\bfr')\varphi_r(\bfr)\varphi_s(\bfr')/\abs{\bfr-\bfr'}$
are the one-electron (kinetic and nuclear potential energies) and two-electron repulsion integrals, respectively.  
The DOCI Hamiltonian,
that we refer to as seniority-zero Hamiltonian in the following, 
is obtained simply by removing all but the terms that preserve the seniority-zero character of
the trial wave function in Eq.~(\ref{eq:S0_parameterization}). As a
result, the one-electron Hamiltonian reduces to its diagonal part while 
only two orbitals (that may be
identical) should remain in the two-electron
repulsion operator:
\begin{eqnarray}\label{eq:HS0}
\hat{H}\rightarrow \hat{H}^{\mathcal{S}_0} &= \displaystyle \sum_{p}h_{pp}\hat{n}_p
+\hat{W}^{\mathcal{S}_0}_{\rm ee},
\end{eqnarray}
where 
\be
\begin{split}\label{eq:WS0}
\hat{W}^{\mathcal{S}_0}_{\rm ee}&=
\sum_{p > q} \psh{pq}{pq} \hat{n}_p \hat{n}_q
 - \sum_{p > q} \sum_\sigma \psh{pq}{qp} \hat{n}_{p\sigma} \hat{n}_{q\sigma} 
\\
 &\quad+ \sum_{pq} \psh{pp}{qq} \hat{P}_p^\dagger \hat{P}_q,
\end{split}
\ee
$\hat{n}_p = \hat{n}_{p\uparrow} + \hat{n}_{p\downarrow}$,
and
$\hat{P}^\dagger_p=\hat{c}_{p\uparrow}^\dagger \hat{c}_{p\downarrow}^\dagger$ is the creation operator of
a pair of electrons in the $p$th orbital. Interestingly, we recognize in
the seniority-zero interaction operator the direct  
$\mathcal{J}_{pq}=\psh{pq}{pq}$, exchange
$\mathcal{K}_{pq}=\psh{pq}{qp}$, and exchange-time-inversion
$\mathcal{L}_{pq}=\psh{pp}{qq}$ integrals which are central in the implementation
of the Piris natural orbital
functionals~\cite{piris2014interacting,piris2017global}. Note that, in
the present work, we use real algebra and therefore
$\mathcal{L}_{pq}=\mathcal{K}_{pq}$. In addition, unlike in Ref.~\onlinecite{limacher2016new},
we do not proceed in Eq.~(\ref{eq:WS0}) with the simplication
$\hat{n}_{p\uparrow}\equiv\hat{n}_{p\downarrow}\equiv\hat{n}_p/2$, which
holds on the seniority-zero
subspace. This is only for convenience, as we will 
implement later on an adiabatic
connection path between the (approximate) seniority-zero and exact
(higher-seniority) Hamiltonians [see Sec.~\ref{subsec:AC}].\\

For two-electron systems in the ground state,
$\hat{H}^{\mathcal{S}_0}$ becomes exact 
if written in the exact natural orbital basis. In this particular
case, the seniority-zero wave
function is the L\"{o}wdin--Shull one~\cite{lowdin1956natural,Mentel2014_JCP}.      
In the general case, $\hat{H}^{\mathcal{S}_0}$ is an approximation to the true
Hamiltonian $\hat{H}$. Nevertheless, the seniority-zero picture may become {\it exact}, in the
spirit of KS-DFT, if it reproduces the true physical (with all higher
seniorities included) 1RDM. As readily seen from Eq.~(\ref{eq:diag_S0_RDM}), it immediately
implies that the seniority-zero and true natural orbitals should match.
In addition, the to-be-determined fictitious seniority-zero wave function $\Psi^{\mathcal{S}_0}$ should reproduce the 
natural orbital occupations of the true physical ground-state wave function
$\Psi_0$:
\be\label{eq:NOO_constraint_general_idea}
n_p^{\Psi^{\mathcal{S}_0}}=n^{\Psi_0}_p.
\ee 
By analogy with DFT for lattice models~\cite{DFT_ModelHamiltonians}, we can
impose the orbital occupation constraint of
Eq.~(\ref{eq:NOO_constraint_general_idea}) [which can be seen as a density constraint in the natural
orbital space] by adjusting the
one-electron energies,
\be
h_{pp}\rightarrow h_{pp}+\overline{\varepsilon}_p,
\ee   
thus leading to the following seniority-zero Schr\"{o}dinger-like
equation:
\be\label{eq:S0_schrodi_eq}
\left[\sum_{p}\left(h_{pp}+\overline{\varepsilon}_p\right)\hat{n}_p
+\hat{W}^{\mathcal{S}_0}_{\rm ee}\right]\ket{\Psi^{\mathcal{S}_0}}=\mathcal{E}^{\mathcal{S}_0}
\ket{\Psi^{\mathcal{S}_0}},
\ee
where $\left\{h_{pp}+\overline{\varepsilon}_p\right\}$, which is unique up to a
constant, can be seen as
the analog (in the natural orbital space) of the KS potential for
seniority-zero wave functions. Note that the extension of the
Hohenberg--Kohn theorem to lattice Hamiltonians
holds also in this
context (see Appendix~\ref{app:HKtheo}). If we compare with the KS Hamiltonian for Hubbard~\cite{DFT_ModelHamiltonians}, the one-electron hopping operator (which is a model for
the kinetic energy operator) is now replaced by the two-electron
seniority-zero repulsion operator $\hat{W}^{\mathcal{S}_0}_{\rm ee}$.
The annihilation-creation of electron pairs
[third term on the right-hand side of Eq.~(\ref{eq:WS0})] delocalizes
the electrons in the natural orbital space, thus ensuring that
all DOCI coefficients are in principle non-zero. This
observation becomes central when establishing a one-to-one correspondence
between 
$\left\{h_{pp}+\overline{\varepsilon}_p\right\}$, that we simply
refer to as {\it potential} in the following, and the ground-state
seniority-zero wave function.\\

In the rest of this work we will show how the
natural orbitals and the potential 
$\left\{\overline{\varepsilon}_p\right\}$ 
can be determined, in principle exactly,
from a universal density matrix functional. This functional enables the
exact evaluation of the true ground-state energy from the fictitious
seniority-zero wave function $\Psi^{\mathcal{S}_0}$.

\subsection{Natural orbital functional theory and orbital rotations}
In RDMFT~\cite{gilbert1975hohenberg}, the exact
ground-state energy can in principle be obtained variationally as
follows,
\begin{eqnarray}\label{eq:VP_dm}
E=\underset{{\bm D}}{\rm min} \Big\{ 
{\rm Tr}\left[{\bm h}{\bm D}\right]+W({\bm D})
\Big\},
\end{eqnarray}
where ${\rm Tr}$ denotes the trace, ${\bm h}\equiv \{h_{pq}\}$ is the
one-electron (kinetic and nuclear potential) Hamiltonian matrix, 
and ${\bm D}\equiv
\{D_{pq}\}$ is a trial (spin-summed) $N$-representable 1RDM. For simplicity,
1RDMs will simply be referred to as {\it density matrices} in the following. 
Within Levy's constrained-search
formalism~\cite{levy1979universal}, the interaction
energy functional reads
\begin{eqnarray}\label{eq:Wgamma}
W({\bm D})=
\underset{\Psi\rightarrow {\bm D}}{\rm min}
\langle
\Psi\vert 
\hat{W}_{\rm ee}
\vert\Psi\rangle.
\end{eqnarray}
Like in {\it natural orbital functional theory} (NOFT)~\cite{piris2014perspective}, we adopt the 
natural orbital representation of the density matrix in the rest of this work. Moreover, for
convenience, we will use an exponential parameterization of the
natural orbitals, like in the MCSCF method~\cite{helgaker2014molecular}. Let $\left\{\varphi_p(\bfr)\right\}$ be an arbitrary (orthonormal) molecular orbital basis. Any natural orbital basis can be obtained from the latter by rotation,
\be\label{eq:orb_rot}
\varphi_p(\bfr)\rightarrow \varphi_{p({\bm \kappa})}(\bfr)=\sum_q\left[e^{-{\bm \kappa}}\right]_{qp}\varphi_q(\bfr),
\ee    
where ${\bm \kappa}=-{\bm
\kappa}^\dagger\equiv\left\{\kappa_{pq}\right\}_{p>q}$. Note that, in
second quantization, Eq.~(\ref{eq:orb_rot}) can be reformulated as
follows~\cite{helgaker2014molecular}, 
\be\label{eq:second_quant_orb_rot}
\hat{c}^\dagger_{p({\bm \kappa})\sigma}=e^{-\hat{\kappa}}\hat{c}^\dagger_{p\sigma}e^{\hat{\kappa}},
\ee
where
$\hat{\kappa}=\sum_{p>q}\kappa_{pq}\sum_\sigma\left(\hat{c}^\dagger_{p\sigma}\hat{c}_{q\sigma}-\hat{c}^\dagger_{q\sigma}\hat{c}_{p\sigma}\right)$
is the real antihermitian singlet rotation operator. The interaction functional in Eq.~(\ref{eq:Wgamma}) can now be rewritten as follows:
\be\label{eq:from_DM_to_nat_orb}
\begin{split}
W\Big({\bm D}\equiv({\bm \kappa},\bfn)\Big)
&=\min_{\Psi\rightarrow ({\bm \kappa},\bfn)}\bra{\Psi}\hat{W}_{\rm ee}\ket{\Psi}
\\
&=\expval{\hat{W}_{\rm ee}}_{\Psi({\bm \kappa},\bfn)},
\end{split}
\ee
where $\bfn\equiv \left\{n_p\right\}$ are the natural orbital
occupancies and the constraint $\Psi\rightarrow ({\bm \kappa},\bfn)$,
which reads 
\be\label{eq:NOO_constraint}
\begin{split}
&\sum_{\sigma}\bra{\Psi}\hat{c}^\dagger_{p({\bm
\kappa})\sigma}\hat{c}_{q({\bm \kappa})\sigma}\ket{\Psi}= n_p\delta_{p({\bm \kappa})q({\bm \kappa})}
\\
&\overset{notation}{\Leftrightarrow} \Psi\underset{\left\{\varphi_{p({\bm
\kappa})}\right\}}{\longrightarrow} {\rm diag}[\bfn],
\end{split}
\ee
means that the rotated orbitals $\left\{\varphi_{p({\bm
\kappa})}\right\}$ are the natural orbitals of $\Psi$ (with
occupations $\bfn$).
In the light of Eq.~(\ref{eq:NOO_constraint}), it seems relevant (but it
is
not compulsory) to expand the minimizing wave
function of Eq.~(\ref{eq:from_DM_to_nat_orb}) in the rotated orbital
basis:
\be\label{eq:Psi_kappa_n_exp_in_rot_MOs}
\ket{\Psi({\bm \kappa},\bfn)}\equiv \sum_I C_I({\bm \kappa},\bfn)
\ket{\Phi_I({\bm \kappa})},
\ee
where $\ket{\Phi_I({\bm \kappa})}$ is a ``rotated'' Slater determinant (\ie, it is constructed
from the rotated orbitals $\left\{\varphi_{p({\bm
\kappa})}\right\}$). Note that, according to
Eq.~(\ref{eq:second_quant_orb_rot}), rotated determinants are connected
to the unrotated ones $\Phi_I\equiv {\Phi_I({\bm \kappa}=0)}$ as
follows, 
\be\label{eq:rot_det_from_unrot_det}
\ket{\Phi_I({\bm \kappa})}=e^{-\hat{\kappa}}\ket{\Phi_I}.
\ee
We should also stress that, unlike its
seniority-zero approximation, the full two-electron repulsion operator
in Eq.~(\ref{eq:full_2e_repulsion_unrotated_basis}) is invariant under
orbital rotation, \ie,
\be\label{eq:full_2e_int_op_rotated_basis}
2\hat{W}_{\rm ee}=
\sum_{pqrs}\psh{pq}{
rs}^{\bm \kappa}
\sum_{\sigma\sigma'}\hat{c}^\dagger_{p({\bm \kappa})\sigma}\hat{c}^\dagger_{q({\bm \kappa})\sigma'}\hat{c}_{s({\bm \kappa})\sigma'}\hat{c}_{r({\bm \kappa})\sigma},
\ee 
where $\psh{pq}{
rs}^{\bm \kappa}=\psh{p({\bm \kappa})q({\bm \kappa})}{
r({\bm \kappa})s({\bm \kappa})}$ are the two-electron integrals
evaluated from the rotated orbitals.\\

For reasons that will become clearer in the
following, we now proceed with a {\it picture change} in
Eq.~(\ref{eq:from_DM_to_nat_orb}):
\be\label{eq:pict_change_wf}
\Psi&{\rightarrow}&\tilde{\Psi}=e^{+\hat{\kappa}}\Psi, 
\ee
where the rotation operator $\hat{\kappa}$ is constructed from
the ${\bm\kappa}$ matrix [see below Eq.~(\ref{eq:second_quant_orb_rot})] which is {\it fixed} when evaluating the interaction
functional of Eq.~(\ref{eq:from_DM_to_nat_orb}). 
When applied to the minimizing wave function, the picture change gives [see
Eq.~(\ref{eq:rot_det_from_unrot_det})]
\be
e^{+\hat{\kappa}}\ket{\Psi({\bm \kappa},\bfn)}\equiv \sum_I C_I({\bm \kappa},\bfn)
\ket{\Phi_I}.
\ee
Therefore, in the new picture, the correct CI coefficients $C_I({\bm
\kappa},\bfn)$ are used in the
(incorrect) unrotated natural orbital basis. The formal advantage is
that trial wave functions can now be expanded 
in the (arbitrary and) fixed orbital basis $\left\{\varphi_p\right\}$.
Obviously, we now need to employ picture-changed operators. For example,
\be\label{eq:exp_val_Wee_picture_change}
\mel{\Psi}{\hat{W}_{\rm ee}}{\Psi}=\mel{\tilde{\Psi}}{\hat{W}_{\rm
ee}({\bm \kappa})}{\tilde{\Psi}},
\ee
where the full (natural orbital-dependent) picture-changed 
two-electron interaction operator $\hat{W}_{\rm
ee}({\bm \kappa})$ reads, according to
Eqs.~(\ref{eq:second_quant_orb_rot}) and (\ref{eq:pict_change_wf}), and
the expression of $\hat{W}_{\rm ee}$ in
Eq.~(\ref{eq:full_2e_int_op_rotated_basis}),
\begin{eqnarray}\label{eq:Weekappa}
\hat{W}_{\rm ee}({\bm \kappa})&=&e^{+\hat{\kappa}}\,\hat{W}_{\rm
ee}\,e^{-\hat{\kappa}} \nonumber \\
&
=&\dfrac{1}{2}\sum_{pqrs}\psh{pq}{
rs}^{\bm
\kappa}\sum_{\sigma\sigma'}\hat{c}^\dagger_{p\sigma}\hat{c}^\dagger_{q\sigma'}\hat{c}_{s\sigma'}\hat{c}_{r\sigma}.
\end{eqnarray}
Note that $\hat{W}_{\rm ee}({\bm \kappa})$ is expressed in terms of the
unrotated creation/annihilation operators because, according to
Eq.~(\ref{eq:second_quant_orb_rot}),
$e^{+\hat{\kappa}}\;\hat{c}^\dagger_{p({\bm
\kappa})\sigma}\;e^{-\hat{\kappa}}=\hat{c}^\dagger_{p\sigma}$.
Finally, since
\be
\delta_{p({\bm \kappa})q({\bm \kappa})}=\bra{\varphi_{p({\bm
\kappa})}}\ket{\varphi_{q({\bm \kappa})}}
=\bra{\varphi_p}\ket{\varphi_q}=\delta_{pq},
\ee 
according to Eq.~(\ref{eq:orb_rot}), the constraint in Eq.~(\ref{eq:NOO_constraint})
becomes in the new picture [see Eqs.~(\ref{eq:second_quant_orb_rot}) and (\ref{eq:pict_change_wf})],
\be\label{eq:picture_change_NOF_constraint}
\begin{split}
&\sum_{\sigma}\bra{\tilde{\Psi}}\hat{c}^\dagger_{p\sigma}\hat{c}_{q\sigma}\ket{\tilde{\Psi}}=n_p\delta_{pq}
\\
&\overset{notation}{\Leftrightarrow}
\tilde{\Psi}\,\rightarrow\,
{\rm diag}[\bfn],
\end{split}
\ee
which means that the unrotated orbitals should be the natural orbitals of
the trial wave function $\tilde{\Psi}$ (with
occupations $\bfn$).\\

To conclude, if we employ the picture change of
Eq.~(\ref{eq:pict_change_wf}), the universal interaction density matrix
functional of Eq.~(\ref{eq:from_DM_to_nat_orb}) can be rewritten as
follows [see Eqs.~(\ref{eq:exp_val_Wee_picture_change}) and
(\ref{eq:picture_change_NOF_constraint})],
\be\label{eq:univ_int_func_nat_orb_rot}
W({\bm \kappa},\bfn)=\min_{\tilde{\Psi}\,\rightarrow\,
{\rm diag}[\bfn]}\bra{\tilde{\Psi}}
\hat{W}_{\rm ee}({\bm \kappa})
\ket{\tilde{\Psi}}.
\ee
We will use the above expression for introducing
seniority-zero wave functions in RDMFT.  

\subsection{Seniority-zero interaction density matrix functional}

On the basis of Eq.~(\ref{eq:univ_int_func_nat_orb_rot}), we define the seniority-zero analog of the interaction
functional as follows,
\be\label{eq:S0_Levy_func_final}
W^{\mathcal{S}_0}({\bm \kappa},\bfn)=\min_{
\tilde{\Psi}\,\rightarrow\,
{\rm diag}[\bfn]
}\bra{\tilde{\Psi}}
\hat{W}^{\mathcal{S}_0}_{\rm ee}({\bm \kappa})
\ket{\tilde{\Psi}},
\ee 
where, in analogy with Eq.~(\ref{eq:WS0}), all but the seniority-zero components
have been removed from the two-electron repulsion operator $\hat{W}_{\rm
ee}({\bm \kappa})$:
\be\label{kappa_dependent_2e_rep}
\begin{split}
\hat{W}^{\mathcal{S}_0}_{\rm ee}({\bm \kappa})
&=
\sum_{p > q} \psh{pq}{pq}^{\bm\kappa} \hat{n}_p \hat{n}_q
 - \sum_{p > q} \sum_\sigma \psh{pq}{qp}^{\bm\kappa} \hat{n}_{p\sigma} \hat{n}_{q\sigma}
\\
 &\quad+ \sum_{pq} \psh{pp}{qq}^{\bm\kappa} \hat{P}_p^\dagger \hat{P}_q.
\end{split}
\ee
Let us now introduce, in the light of Sec.~\ref{subsec:seniority_zero},
the ground-state wave function $\Psi^{\mathcal{S}_0}({\bm\kappa},\bfn)$ of the
seniority-zero Hamiltonian $\hat{W}^{\mathcal{S}_0}_{\rm ee}({\bm
\kappa})+
\sum_p\varepsilon_p({\bm\kappa},\bfn)\hat{n}_p$ with natural orbital occupations
$\bfn$. For {\it any} wave function $\Psi$ fulfilling the (not
necessarily natural) orbital occupation constraint
\be\label{eq:orb_occ_constraint}
\left\{\bra{\Psi}\hat{n}_p\ket{\Psi}\right\}=\bfn,
\ee
that we simply denote $\Psi\rightarrow
\bfn$, we have, according
to the variational principle,
\begin{eqnarray}\label{eq:VP_wS0_kappa}
\begin{split}
&\bra{\Psi^{\mathcal{S}_0}({\bm\kappa},\bfn) }\hat{W}^{\mathcal{S}_0}_{\rm ee}({\bm
\kappa})+\sum_p\varepsilon_p({\bm\kappa},\bfn)\hat{n}_p\ket{\Psi^{\mathcal{S}_0}({\bm\kappa},\bfn)} 
\\
&\leq \bra{ \Psi } \hat{W}^{\mathcal{S}_0}_{\rm ee}({\bm
\kappa})+\sum_p\varepsilon_p({\bm\kappa},\bfn)\hat{n}_p\ket{\Psi}
,
\end{split}
\end{eqnarray}
thus leading to
\be
\bra{\Psi^{\mathcal{S}_0}({\bm\kappa},\bfn)}\hat{W}^{\mathcal{S}_0}_{\rm ee}({\bm
\kappa})\ket{\Psi^{\mathcal{S}_0}({\bm\kappa},\bfn)}\underset{\Psi\rightarrow
\bfn}{\leq}
\bra{\Psi}\hat{W}^{\mathcal{S}_0}_{\rm ee}({\bm
\kappa})\ket{\Psi}.
\nonumber\\
\ee
As a result, the density matrix constraint in the minimization of
Eq.~(\ref{eq:S0_Levy_func_final}), which is obvioulsy fulfilled by $\Psi^{\mathcal{S}_0}({\bm\kappa},\bfn)$, can be relaxed, thus becoming the
orbital occupation constraint of Eq.~(\ref{eq:orb_occ_constraint}). This
leads to the
final Levy's constrained-search expression for the seniority-zero interaction energy functional:    
\begin{eqnarray}\label{eq:minPsi_WSOee}
W^{\mathcal{S}_0}({\bm \kappa},\bfn)&=&
\underset{\Psi\rightarrow \bfn}{\rm min}
\bra{\Psi}\hat{W}^{\mathcal{S}_0}_{\rm ee}({\bm
\kappa})\ket{\Psi}
\\
\label{eq:minPsi_WSOee_reached}
&=&
\bra{\Psi^{\mathcal{S}_0}({\bm\kappa},\bfn)}\hat{W}^{\mathcal{S}_0}_{\rm ee}({\bm
\kappa})\ket{\Psi^{\mathcal{S}_0}({\bm\kappa},\bfn)}.
\end{eqnarray}
As the orbital occupation constraint is analogous to a density
constraint on a lattice~\cite{DFT_ModelHamiltonians,senjean2018site}, we can use convenient tools from DFT
such as the Lieb maximization~\cite{LFTransform-Lieb,senjean2018site} in order to evaluate the functional.
Indeed, for {\it any} ``potential'' ${\bm\varepsilon}\equiv\left\{\varepsilon_p\right\}$,
applying the variational principle to the seniority-zero Hamiltonian $\hat{W}^{\mathcal{S}_0}_{\rm ee}({\bm
\kappa})+
\sum_p\varepsilon_p\hat{n}_p$ with ground-state energy
$\mathcal{E}^{\mathcal{S}_0}({\bm
\kappa},{\bm\varepsilon})$ gives
\be
\begin{split}
&\mathcal{E}^{\mathcal{S}_0}({\bm
\kappa},{\bm\varepsilon})
\\
&\leq
\bra{\Psi^{\mathcal{S}_0}({\bm\kappa},\bfn) }\hat{W}^{\mathcal{S}_0}_{\rm ee}({\bm
\kappa})+\sum_p\varepsilon_p\hat{n}_p\ket{\Psi^{\mathcal{S}_0}({\bm\kappa},\bfn)},   
\end{split}
\ee
thus leading to the alternative expression [see Eqs.~(\ref{eq:orb_occ_constraint}) and (\ref{eq:minPsi_WSOee_reached})]
\be\label{eq:S0fun_Lieb_max}
W^{\mathcal{S}_0}({\bm \kappa},\bfn)=\max_{{\bm\varepsilon}}
\left\{\mathcal{E}^{\mathcal{S}_0}({\bm
\kappa},{\bm\varepsilon})-({\bm\varepsilon}\vert \bfn)\right\},
\ee  
where $({\bm\varepsilon}\vert \bfn)=\sum_p\varepsilon_pn_p$. 
The practical implementation of Eq.~(\ref{eq:S0fun_Lieb_max}) will be
discussed further in Secs.~\ref{sec:comput_details} and
\ref{sec:discussion}.\\

Finally, we would like to emphasize that, even though the orbital occupation
constraint of Eq.~(\ref{eq:orb_occ_constraint}) is written, for
convenience, in an arbitrary and fixed 
(unrotated) orbital basis, the seniority-zero interaction functional is
a {\it universal} functional of the density matrix. In other words,
unlike in DFT for lattice Hamiltonians~\cite{DFT_ModelHamiltonians}, the (natural, in the present
context) orbital basis on which the theory relies (through its
dependence in ${\bm \kappa}$) is an
additional basic variable that has to be optimized for a given system, exactly like in
NOFT. This point can be further highlighted by making the dependence
of the functional on the natural orbitals more explicit. Indeed, since
$e^{-\hat{\kappa}}\Psi^{\mathcal{S}_0}({\bm\kappa},\bfn)$
is the ground state of [see Eqs.~(\ref{eq:second_quant_orb_rot}) and (\ref{eq:VP_wS0_kappa})]
\be\label{eq:rotated_S0_hamilt}
\begin{split}
&e^{-\hat{\kappa}}\left(\hat{W}^{\mathcal{S}_0}_{\rm ee}({\bm
\kappa})+
\sum_p\varepsilon_p({\bm\kappa},\bfn)\hat{n}_p\right)e^{\hat{\kappa}}
\\
&=
\sum_{p > q} \psh{pq}{pq}^{\bm\kappa} \hat{n}_{p({\bm\kappa})}
\hat{n}_{q({\bm\kappa})}
\\
&\quad - \sum_{p > q} \sum_\sigma \psh{pq}{qp}^{\bm\kappa} \hat{n}_{p({\bm\kappa})\sigma} \hat{n}_{q({\bm\kappa})\sigma}
\\
 &\quad+ \sum_{pq} \psh{pp}{qq}^{\bm\kappa} \hat{P}_{p({\bm\kappa})}^\dagger
\hat{P}_{q({\bm\kappa})}+\sum_p\varepsilon_p({\bm\kappa},\bfn)\hat{n}_{p({\bm\kappa})}
,
\end{split}
\ee
with occupation numbers in the rotated basis [see Eq.~(\ref{eq:orb_occ_constraint})]
\be
\begin{split}
&\left\{\bra{e^{-\hat{\kappa}}\Psi^{\mathcal{S}_0}({\bm\kappa},\bfn)}
\hat{n}_{p({\bm\kappa})}
\ket{e^{-\hat{\kappa}}\Psi^{\mathcal{S}_0}({\bm\kappa},\bfn)}
\right\}
\\
&=
\left\{
\bra{\Psi^{\mathcal{S}_0}({\bm\kappa},\bfn)}
\hat{n}_{p}
\ket{\Psi^{\mathcal{S}_0}({\bm\kappa},\bfn)}
\right\}
\\
&\equiv \bfn,
\end{split}
\ee
it may be denoted as 
\be
e^{-\hat{\kappa}}\Psi^{\mathcal{S}_0}({\bm\kappa},\bfn)
\equiv 
\overline{\Psi}^{\mathcal{S}_0}\left(\left\{\varphi_{p({\bm
\kappa})}\right\},\bfn\right)
\ee
and described as the ({\it ab initio}) seniority-zero analog of the
ground-state wave function, both being written in the same basis of natural orbitals $\left\{\varphi_{p({\bm
\kappa})}\right\}$ with occupations $\bfn$. As the seniority-zero
Hamiltonian of 
Eq.~(\ref{eq:rotated_S0_hamilt}) is a universal functional of the
density matrix, so are $\overline{\Psi}^{\mathcal{S}_0}\left(\left\{\varphi_{p({\bm
\kappa})}\right\},\bfn\right)$ and the resulting seniority-zero
interaction functional [see Eq.~(\ref{eq:minPsi_WSOee_reached})]:   
\be\label{eq:simpler_exp_S0_func}
&&W^{\mathcal{S}_0}({\bm \kappa},\bfn)
\nonumber
\\
&&
=
\bra{\overline{\Psi}^{\mathcal{S}_0}\left(\left\{\varphi_{p({\bm
\kappa})}\right\},\bfn\right)}
e^{-\hat{\kappa}}\hat{W}^{\mathcal{S}_0}_{\rm ee}({\bm
\kappa})e^{\hat{\kappa}}
\ket{\overline{\Psi}^{\mathcal{S}_0}\left(\left\{\varphi_{p({\bm
\kappa})}\right\},\bfn\right)}
\nonumber
\\
&&=
\bra{\overline{\Psi}^{\mathcal{S}_0}\left(\left\{\varphi_{p({\bm
\kappa})}\right\},\bfn\right)}
\hat{W}_{\rm ee}
\ket{\overline{\Psi}^{\mathcal{S}_0}\left(\left\{\varphi_{p({\bm
\kappa})}\right\},\bfn\right)}
\nonumber
\\
&&\equiv
W^{\mathcal{S}_0}(\left\{\varphi_{p({\bm \kappa})}\right\},\bfn).
\ee   
For practical purposes, the (less
intuitive) expressions in Eqs.~(\ref{eq:minPsi_WSOee}) and
(\ref{eq:S0fun_Lieb_max}) are much more appealing as they do not
require the implementation of a specific (seniority-zero in this context) wave
function ansatz. Indeed, starting from a regular (seniority-zero) HF
calculation, the functional can be evaluated by setting to zero, within the post-HF
treatment, the one- and
two-electron integrals that do not preserve seniority. The dependence in
${\bm \kappa}$ of the remaining integrals [see
Eq.~(\ref{kappa_dependent_2e_rep})] enables the optimization of the
natural orbitals, as further discussed in the next section.

\subsection{Exact self-consistent seniority-zero equations}

Let us start from the regular NOFT~\cite{piris2014perspective} variational energy expression that we simply rewrite as follows
[see Eqs.~(\ref{eq:VP_dm}) and (\ref{eq:from_DM_to_nat_orb})]:
\begin{eqnarray}\label{eq:E0_var_kappa_n}
E= \min_{\bm{\kappa},\bfn} \Big \lbrace
(\bm{h}(\bm{\kappa})|\bfn)
+
W(\bm{\kappa},\bfn) 
 \Big \rbrace,
\end{eqnarray}
where 
$(\bm{h}(\bm{\kappa})|\bfn) = \sum_p h_{pp}(\bm{\kappa}) n_p$ and
\be\label{eq:kappa_dependent_1int}
h_{pp}(\bm{\kappa})=\bra{\varphi_{p(\bm{\kappa})}}\hat{h}\ket{\varphi_{p(\bm{\kappa})}}.
\ee
If we
now split the universal interaction functional into
seniority-zero and complementary (higher-seniority) parts,
\be\label{eq:seniority_separation}
\begin{split}
W(\bm{\kappa},\bfn)&=W^{\mathcal{S}_0}(\bm{\kappa},\bfn)+\left(W(\bm{\kappa},\bfn)-W^{\mathcal{S}_0}(\bm{\kappa},\bfn)\right)  
\\
&=W^{\mathcal{S}_0}(\bm{\kappa},\bfn)+\overline{W}^{\mathcal{S}}(\bm{\kappa},\bfn),
\end{split}
\ee 
the exact ground-state energy can be rewritten [see Eq.~(\ref{eq:minPsi_WSOee})] as follows: 
\begin{eqnarray}
E=
\underset{{\bm \kappa},\bfn}{\rm min} \Big \lbrace
\min_{\Psi \rightarrow \bfn} \Big \lbrace
&&
\left({\bm h}({\bm
\kappa})\middle\vert \bfn^\Psi\right)
+
\bra{\Psi}\hat{W}^{\mathcal{S}_0}_{\rm ee}({\bm
\kappa})\ket{\Psi} 
\nonumber \\
&&+
 \overline{W}^{\mathcal{S}}(\bm{\kappa},\bfn^\Psi)
\Big \rbrace 
\Big\rbrace,
\end{eqnarray}
thus leading to
\begin{eqnarray}\label{eq:double_min_var_ener_exp}
E=
\underset{{\bm \kappa}}{\rm min} \Big \lbrace
\min_{\Psi} \Big \lbrace
&&
\left({\bm h}({\bm
\kappa})\middle\vert \bfn^\Psi\right)
+
\bra{\Psi}\hat{W}^{\mathcal{S}_0}_{\rm ee}({\bm
\kappa})\ket{\Psi} 
\nonumber \\
&&+
 \overline{W}^{\mathcal{S}}(\bm{\kappa},\bfn^\Psi)
\Big \rbrace 
\Big\rbrace,
\end{eqnarray}
or, equivalently,
\be\label{eq:true_final_var_ener_exp}
E=
\min_{{\bm \kappa},\Psi} \left\{
\left
\langle\sum_ph_{pp}({\bm\kappa})\hat{n}_p+\hat{W}^{\mathcal{S}_0}_{\rm ee}({\bm
\kappa})\right\rangle_{\Psi} 
+\overline{W}^{\mathcal{S}}(\bm{\kappa},\bfn^\Psi)
\right\} 
\nonumber
\\
\ee
where $\bfn^{\Psi}\equiv \left\{\bra{\Psi}\hat{n}_p\ket{\Psi}\right\}$.
We stress again that, unlike in DFT for
lattices~\cite{DFT_ModelHamiltonians,senjean2018site}, the energy is
obtained in this context by means of a {\it double} minimization. The
first one (with respect to the many-body
wave function $\Psi$) aims at reproducing the natural orbital
occupancies. The second one (with respect to ${\bm\kappa}$), where
the one- and two-electron integrals vary [see
Eqs.~(\ref{eq:kappa_dependent_1int}) and (\ref{kappa_dependent_2e_rep}), respectively], enables the variational
calculation of the natural
orbitals.\\

For convenience, we choose the initial (unrotated) orbital basis to be
the exact natural orbital basis of the system under study. In other
words, the minimum in Eq.~(\ref{eq:true_final_var_ener_exp}) is reached
when ${\bm\kappa}=0$. By denoting
\be\label{eq:S0_dens_fun_notation}
\overline{W}^{\mathcal{S}}(\bfn)\equiv \overline{W}^{\mathcal{S}}(\bm{\kappa}=0,\bfn),
\ee
we can reformulate the variational principle as follows:
\be\label{eq:min_ener_psi_kappa_zero}
\begin{split}
E&=\min_{\Psi}\left\{\bra{\Psi}\sum_ph_{pp}\hat{n}_p+\hat{W}^{\mathcal{S}_0}_{\rm ee}
\ket{\Psi}+\overline{W}^{\mathcal{S}}(\bfn^\Psi)\right\}
\\
&=\sum_ph_{pp}n^{\Psi^{\mathcal{S}_0}}_p+\bra{\Psi^{\mathcal{S}_0}}\hat{W}^{\mathcal{S}_0}_{\rm ee}
\ket{\Psi^{\mathcal{S}_0}}+\overline{W}^{\mathcal{S}}(\bfn^{\Psi^{\mathcal{S}_0}}).
\end{split}
\ee
From the corresponding Euler--Lagrange equation, we conclude that the
minimizing wave function $\Psi^{\mathcal{S}_0}$, which reproduces the
exact natural orbital occupancies, is the ground-state solution to the
following
self-consistent and seniority-zero Schr\"{o}dinger-like equation: 
\be\label{eq:final_sc_S0_eq}
\hat{\mathcal{H}}^{\mathcal{S}_0}({\bfn}^{\Psi^{\mathcal{S}_0}})\ket{\Psi^{\mathcal{S}_0}}=\mathcal{E}^{\mathcal{S}_0}\ket{\Psi^{\mathcal{S}_0}},
\ee
where 
\be\label{eq:final_sc_S0_eq_hamil}
\hat{\mathcal{H}}^{\mathcal{S}_0}(\bfn)=\sum_p\left(h_{pp}+\dfrac{\partial
\overline{W}^{\mathcal{S}}(\bfn)}{\partial
n_p}\right)\hat{n}_p+\hat{W}^{\mathcal{S}_0}_{\rm ee}.
\ee
Note that the lower bound in Eq.~(\ref{eq:min_ener_psi_kappa_zero}),
which is the exact ground-state energy,
can be rewritten as follows, 
\be\label{eq:ener_from_expvalH}
E=\expval{\hat{H}}_{\Psi^{\mathcal{S}_0}}+\overline{W}^{\mathcal{S}}(\bfn^{\Psi^{\mathcal{S}_0}}),
\ee
since $\Psi^{\mathcal{S}_0}$ is a seniority-zero wave
function.\\

Interestingly, we conclude from Eq.~(\ref{eq:final_sc_S0_eq_hamil}) that the diagonal one-electron integral correction
introduced in Eq.~(\ref{eq:S0_schrodi_eq}), for the purpose of
reproducing the exact natural orbital occupancies, is a density matrix
functional quantity which can be evaluated as follows:
\be
\overline{\varepsilon}_p=\left.\dfrac{\partial
\overline{W}^{\mathcal{S}}(\bfn)}{\partial
n_p}\right|_{\bfn={\bfn}^{\Psi^{\mathcal{S}_0}}}.
\ee
Further physical insight can be obtained from the seniority separation
in Eq.~(\ref{eq:seniority_separation}), which can be rewritten as
follows:
\be\label{eq:phys_inter_comp_fun}
\overline{W}^{\mathcal{S}}(\bfn)&=&\bra{\Psi(\bfn)}\hat{W}_{\rm
ee}\ket{\Psi(\bfn)}-\bra{\Psi^{\mathcal{S}_0}(\bfn)}\hat{W}^{\mathcal{S}_0}_{\rm ee}
\ket{\Psi^{\mathcal{S}_0}(\bfn)}
\nonumber
\\
&=&
\bra{\Psi(\bfn)}\hat{W}_{\rm
ee}\ket{\Psi(\bfn)}-\bra{\Psi^{\mathcal{S}_0}(\bfn)}\hat{W}_{\rm ee}
\ket{\Psi^{\mathcal{S}_0}(\bfn)}
\nonumber\\
&=&
\bra{\Psi(\bfn)}\hat{H}
\ket{\Psi(\bfn)}-\bra{\Psi^{\mathcal{S}_0}(\bfn)}\hat{H}
\ket{\Psi^{\mathcal{S}_0}(\bfn)},
\ee
where both physical $\Psi(\bfn)$ and seniority-zero $\Psi^{\mathcal{S}_0}(\bfn)$ wave functions share, in addition
to the occupation numbers $\bfn$, the same $\bfn$-independent natural
orbital basis which, in the present case, is the unrotated orbital
basis. As a result, the ``potential'' $\overline{\varepsilon}_p$
describes the higher-seniority contributions to the energy variation
induced by infinitesimal changes in the natural orbital occupation
$n_p$.\\    

Let us finally emphasize that $\overline{W}^{\mathcal{S}}(\bfn)$ is evaluated with
the {\it exact} natural orbitals. According to Eqs.~(\ref{eq:double_min_var_ener_exp}) and
(\ref{eq:min_ener_psi_kappa_zero}), the latter can be determined from the following stationarity
condition, 
\be\label{eq:station_cond_nat_orb}
\begin{split}
&\Bigg[(\partial \bm{h}(\bm{\kappa})/\partial \kappa_{pq}\vert {\bfn}^{\Psi^{\mathcal{S}_0}})
+\bra{\Psi^{\mathcal{S}_0}}\partial \hat{W}^{\mathcal{S}_0}_{\rm ee}(\bm{\kappa})/\partial \kappa_{pq}
\ket{\Psi^{\mathcal{S}_0}}
\\
&+
\dfrac{\partial \overline{W}^{\mathcal{S}}({\bm \kappa},{\bfn
}^{\Psi^{\mathcal{S}_0}})}{\partial
\kappa_{pq}}
\Bigg]_{{\bm\kappa}=0}=0
,
\end{split}
\ee
which can be rewritten more explicitly as follows [see Appendix~\ref{app:proof}],
\be\label{eq:final_sc_eq_nat_orb}
\begin{split}
&2\mathcal{F}_{pq}\left(n^{\Psi^{\mathcal{S}_0}}_p-n^{\Psi^{\mathcal{S}_0}}_q\right)
+\left\langle
\varphi_q
\middle\vert
\dfrac{\delta \overline{W}^{\mathcal{S}}\left(\left\{\varphi_p\right\},{\bfn
}^{\Psi^{\mathcal{S}_0}}\right)}{\delta \varphi_p}
\right\rangle
\\
&-
\left\langle
\varphi_p
\middle\vert
\dfrac{\delta \overline{W}^{\mathcal{S}}\left(\left\{\varphi_q\right\},{\bfn
}^{\Psi^{\mathcal{S}_0}}\right)}{\delta \varphi_q}
\right\rangle
\\
&+\sum_r\left[2\psh{rp}{rq}-\psh{rp}{qr}\right]
\\
&\times
\left[\expval{\left(\hat{n}_r-\expval{\hat{n}_r}_{\Psi^{\mathcal{S}_0}}\right)
\hat{n}_p}_{\Psi^{\mathcal{S}_0}}
-\expval{
\left(
\hat{n}_r
-\expval{\hat{n}_r}_{\Psi^{\mathcal{S}_0}}
\right)
\hat{n}_q}_{\Psi^{\mathcal{S}_0}}
\right]
\\
&
+4\sum_r\psh{pq}{rr}
\\
&
\times\left[
(1-\delta_{rp})\expval{\hat{P}_r^\dagger
\hat{P}_p}_{\Psi^{\mathcal{S}_0}}
-
(1-\delta_{rq})
\expval{\hat{P}_r^\dagger
\hat{P}_q}_{\Psi^{\mathcal{S}_0}}
\right]=0,
\end{split}
\ee
where $\mathcal{F}_{pq}\equiv \mathcal{F}_{pq}\left(\left\{\varphi_r\right\},{\bfn
}^{\Psi^{\mathcal{S}_0}}\right)$, and 
\be\label{eq:DM_fun_Fock_matrix}
\mathcal{F}_{pq}\left(\left\{\varphi_r\right\},{\bfn
}\right)&=h_{pq}+\dfrac{1}{2}\sum_r\left[2\psh{rp}{rq}-\psh{rp}{qr}\right]n_r \nonumber \\
\ee
is the density matrix functional Fock matrix element. Note that regular HF and NOFT equations can be
recovered from Eq.~(\ref{eq:final_sc_eq_nat_orb}). In both
cases, the two summations over the natural orbitals $r$ should be removed, either
because they are equal to zero (HF case) or because they are incorporated into
the (full) interaction functional ${W}\left(\left\{\varphi_p\right\},{\bfn
}\right)$ [NOFT case]. In the HF case, the
complementary (higher-seniority) density matrix functional $\overline{W}^{\mathcal{S}}\left(\left\{\varphi_p\right\},{\bfn
}\right)$ is neglected
and the usual Fock operator expression for doubly occupied orbitals is
recovered:     
\be
\mathcal{F}_{pq}\overset{\rm
HF}{\longrightarrow}h_{pq}+\sum^{\rm occ.}_r\left[2\psh{rp}{rq}-\psh{rp}{qr}\right],
\ee
thus leading to the HF equations. In the NOFT case, one would proceed
with the following substitutions,
\be
\begin{split}
\mathcal{F}_{pq}&\overset{\rm
NOFT}{\longrightarrow}h_{pq},
\\
\overline{W}^{\mathcal{S}}\left(\left\{\varphi_p\right\},{\bfn
}\right)
&\overset{\rm
NOFT}{\longrightarrow}{W}\left(\left\{\varphi_p\right\},{\bfn
}\right),
\end{split}
\ee
thus leading to the regular NOFT equations for the natural
orbitals~\cite{piris2014perspective}. In the latter case, the occupation numbers are
directly used as basic variables, {\ie} they are not calculated from an
auxiliary many-body wave function, unlike in the present approach. The
construction of density matrix functional approximations to
$\overline{W}^{\mathcal{S}}\left(\left\{\varphi_{p
}\right\},{\bfn
}\right)$ as well as the self-consistent implementation of
Eq.~(\ref{eq:final_sc_eq_nat_orb}), which are essential for turning the
theory into a practical computational method, are left
for future work. In the rest of this work, we
assume that the exact natural orbitals of the system under study are
known and we focus on the evaluation of the complementary (higher-seniority)
natural orbital occupation functional energy
$\overline{W}^{\mathcal{S}}(\bfn^{\Psi^{\mathcal{S}_0}})$ [see
Eq.~(\ref{eq:min_ener_psi_kappa_zero})]. 

\section{Adiabatic connection formalisms}\label{sec:AC_formalism}

We focus in the following on the practical evaluation of the
complementary higher-seniority correlation energy
$\overline{W}^{\mathcal{S}}(\bfn^{\Psi^{\mathcal{S}_0}})$ [see
Eq.~(\ref{eq:ener_from_expvalH})] which has been described in the
previous sections as an implicit functional of the density matrix. For
that purpose, we derive in Sec.~\ref{subsec:AC} an exact adiabatic
connection (AC) formula in the basis of the exact natural orbitals.
Along such a
(so-called constrained) AC path, the occupation of
the (true) natural orbitals is held constant. Approximations along the AC
(based either on second-order perturbation theory or linear
interpolations) are then discussed in Sec.~\ref{subsec:approx_AC}. For analysis purposes, we
finally investigate in Sec.~\ref{sec:relaxedAC} a simpler AC where the natural orbital occupations
constraint is relaxed. In this case, the reference 
seniority-zero wave function does not reproduce the true density matrix
anymore. Nevertheless, integrating over the full relaxed AC path will still
provide the exact ground-state energy. For comparison with the
constrained AC, we also derive a second-order
perturbation theory along the relaxed AC. The performance of the
various approximations will be discussed in the separate Sec.~\ref{sec:discussion}. 

\subsection{Exact constrained adiabatic connection}\label{subsec:AC}

In order to obtain further insight into the higher-seniority natural orbital occupation functional of Eq.~(\ref{eq:S0_dens_fun_notation}), we
propose to construct an AC path between the fictitious
seniority-zero $\Psi^{\mathcal{S}_0}$ wave function and the physical
(fully-interacting) one $\Psi_0$ in
the {\it exact natural orbital basis of the latter} $\left\{\varphi_{p
}\right\}$. For that
purpose, we consider the following partially-interacting (ground-state) Schr\"{o}dinger
equation, 
\be\label{eq:AC_Schrodinger_eq}
\hat{H}^\lambda\ket{\Psi_0^\lambda}=\mathcal{E}_0^\lambda\ket{\Psi_0^\lambda},
\ee
where $\lambda$ varies continuously in the range $0\leq \lambda\leq
1$. Note that we label as ``0'' the ground state along the AC since
the excited states will come into play (later in
Sec.~\ref{subsec:approx_AC}) when solving $\hat{H}^\lambda$ in
perturbation theory.\\

Let us denote 
\be\label{eq:Hamilt_along_AC}
\hat{H}^\lambda\equiv \hat{H}^\lambda\left({\bm
\varepsilon}^\lambda\right)
\ee
and
$\mathcal{E}_0^\lambda\equiv\mathcal{E}_0^\lambda\left({\bm
\varepsilon}^\lambda\right)$, where the partially-interacting
Hamiltonian reads as follows, for an arbitrary potential
$\bepsi\equiv\left\{\varepsilon_p\right\}$, 
\begin{eqnarray}\label{eq:Hlambda_S0}
\hat{H}^\lambda({\bm \varepsilon})
&=&\sum_p \varepsilon_p \hat{n}_p + \hat{W}_{\rm
ee}^{\mathcal{S}_0}+\lambda \hat{\mathcal{V}},
\\
\hat{\mathcal{V}} &=& \hat{W}_{\rm ee} - \hat{W}_{\rm ee}^{\mathcal{S}_0} + \sum_{p\neq q}h_{pq}\hat{n}_{pq},
\end{eqnarray}
and 
$\hat{n}_{pq}=\sum_\sigma\hat{a}^\dagger_{p\sigma}
\hat{a}_{q\sigma}$. The potential ${\bm
\varepsilon}^\lambda\equiv\left\{\varepsilon^\lambda_p\right\}$ is determined along the AC from the natural orbital
occupation constraint
\be\label{eq:AC_occ_constraint}
\left\{\expval{\hat{n}_p}_{\Psi_0^\lambda}\right\}\equiv
\bfn^{\Psi_0^\lambda}=\bfn^{\Psi_0}=\bfn^{\Psi^{\mathcal{S}_0}},
\ee
which, in complete analogy with Eq.~(\ref{eq:S0fun_Lieb_max}), can be
rewritten as a Lieb maximization problem~\cite{LFTransform-Lieb}:
\be\label{eq:AC_Lieb_max}
{\bm
\varepsilon}^\lambda=\arg\max_{{\bm
\varepsilon}}\left\{\mathcal{E}_0^\lambda({\bm
\varepsilon})-\left({\bm\varepsilon}\middle\vert{\bfn}^{\Psi_0}\right)\right\},
\ee
where $\mathcal{E}_0^\lambda({\bm \varepsilon})$ is the ground-state
energy of $\hat{H}^\lambda({\bm\varepsilon})$.\\

As readily seen from Eq.~(\ref{eq:Hlambda_S0}), for $\lambda=1$, the
true physical Hamiltonian is recovered when $\varepsilon^{\lambda=1}_p=h_{pp}$
(the orbital
occupation constraint of Eq.~(\ref{eq:AC_occ_constraint}) is fulfilled
in this case), thus leading to
\be\label{eq:ener_lambda_1}
\mathcal{E}_0^{\lambda=1}=E=\left({\bm\varepsilon}^{\lambda=1}\middle\vert{\bfn}^{\Psi_0}\right)+\expval{\hat{W}_{\rm
ee}}_{\Psi_0},
\ee
where we used the fact that, by construction,
\be
\expval{\hat{n}_{pq}}_{\Psi_0}\overset{p\neq q}{=}0.
\ee
On the other hand, 
the seniority-zero wave function $\Psi^{\mathcal{S}_0}$ with the
same occupation numbers is recovered when $\lambda=0$, thus leading to
the following simplified energy expression,
\be\label{eq:ener_lambda_0}
\mathcal{E}_0^{\lambda=0}=\expval{\hat{W}^{\mathcal{S}_0}_{\rm
ee}}_{\Psi^{\mathcal{S}_0}}+\left({\bm\varepsilon}^{\lambda=0}\middle\vert{\bfn}^{\Psi_0}\right).
\ee 
We stress that, by construction, the density matrix (which is here
evaluated in the exact physical natural orbital basis) is diagonal in both $\lambda=0$ and
$\lambda=1$ limits. 
However, unlike in Ref.~\onlinecite{pernal2018electron}, we do {\it
not}
assume or expect the density matrix to remain strictly diagonal along the AC. 
The present AC relies on an orbital occupation constraint, not on a density matrix
one. Nevertheless, for the systems discussed in
Sec.~\ref{sec:discussion}, we have observed numerically that the off-diagonal elements of the 1RDM do not deviate
substantially from zero when $0<\lambda<1$.\\        

We can now evaluate along the AC path the complementary higher-seniority
correlation energy of Eq.~(\ref{eq:phys_inter_comp_fun}), in the
particular case where $\bfn=\bfn^{\Psi_0}=\bfn^{\Psi^{\mathcal{S}_0}}$. From the original expression 
\be\label{eq:complement_corr_ener_by_difference}
\overline{W}^{\mathcal{S}}\equiv \overline{W}^{\mathcal{S}}(\bfn^{\Psi_0})=
\expval{\hat{W}_{\rm
ee}}_{\Psi_0}-\expval{\hat{W}^{\mathcal{S}_0}_{\rm
ee}}_{\Psi^{\mathcal{S}_0}},
\ee
which can be rewritten as follows [see Eq.~(\ref{eq:ener_lambda_1}) and Eq.~(\ref{eq:ener_lambda_0})],
\be
\begin{split}
\overline{W}^{\mathcal{S}}&=\left[\mathcal{E}_0^{\lambda}-\left({\bm\varepsilon}^{\lambda}\middle\vert\bfn^{\Psi_0}\right)\right]_{\lambda=1}
\\
&\quad-\left[\mathcal{E}_0^{\lambda}-\left({\bm\varepsilon}^{\lambda}\middle\vert\bfn^{\Psi_0}\right)\right]_{\lambda=0},
\end{split}
\ee
we obtain the following exact AC formula, 
\be
\overline{W}^{\mathcal{S}}
&=&\int^1_0d\lambda\left[\dfrac{\partial 
\mathcal{E}_0^\lambda}{\partial \lambda}-\left(\dfrac{\partial{\bm
\varepsilon}^\lambda}{\partial\lambda}\middle\vert\bfn^{\Psi_0}\right)
\right]
\\
\label{eq:final_AC_formula}
&=&
\int^1_0d\lambda\,\overline{\mathcal{W}}^{\mathcal{S},\lambda},
\ee
where, according to the Hellmann--Feynman theorem and the
orbital occupation constraint of Eq.~(\ref{eq:AC_occ_constraint}),
the higher-seniority integrand reads
\be
\label{eq:integrand_mathcalV}
\overline{\mathcal{W}}^{\mathcal{S},\lambda}
&=&\expval{\hat{\mathcal{V}}}_{\Psi_0^\lambda}
\\
\label{eq:corr_int}
&=&\expval{\hat{W}_{\rm
ee}-\hat{W}_{\rm ee}^{\mathcal{S}_0}}_{\Psi_0^\lambda}
+\sum_{p\neq q}
{h_{pq}}\expval{\hat{n}_{pq}}_{\Psi_0^\lambda}
.
\ee
Note that $\overline{\mathcal{W}}^{\mathcal{S},\lambda=0}=0$,
by construction. Moreover, as pointed out previously, 
$\expval{\hat{n}_{pq}}_{\Psi_0^\lambda}$ vanishes in both
fully-interacting ($\lambda=1$) and seniority-zero ($\lambda=0$) limits but
it can in principle deviate from zero when $0 < \lambda < 1$.\\

Once the $\lambda$-dependent integrand is determined, the total
ground-state energy can be
evaluated, in principle exactly, as follows [see
Eq.~(\ref{eq:ener_from_expvalH})], 
\be\label{eq:exact_GS_ener_trueAC}
E=\expval{\hat{H}}_{\Psi^{\mathcal{S}_0}}+\int^1_0d\lambda\,\overline{\mathcal{W}}^{\mathcal{S},\lambda}.
\ee
The above formula and Eq.~(\ref{eq:corr_int}) are interesting in several ways. Firstly, 
as readily seen, the integrand plays a
central role in the description of
higher-seniority energy contributions. Secondly, they can serve
as a basis for the development of approximations, as discussed further
in the following.  

\subsection{Approximations along the adiabatic
connection}\label{subsec:approx_AC}

As done before in the context of DFT [see, for example,
Refs.~\cite{cornaton2013analysis,vuckovic2016exchange}
and the references therein], we can construct various approximate
schemes by simplifying the integrand expression. For example, by analogy
with G\"{o}rling--Levy second-order perturbation theory
(PT2)~\cite{gorling1993correlation,gorling1994exact},     
we may assume that the integrand varies linearly (with the
slope obtained in the seniority-zero $\lambda=0$ limit)
along the AC:
\begin{eqnarray}\label{eq:integrand_PT2}
\overline{\mathcal{W}}^{\mathcal{S},\lambda }
\overset{\rm PT2}{\approx} \lambda \left.\dfrac{\partial
\overline{\mathcal{W}}^{\mathcal{S},\lambda}}{\partial \lambda}
\right|_{\lambda = 0}.
\end{eqnarray}
According to Eq.~(\ref{eq:integrand_mathcalV}) and perturbation theory
that we apply to Eq.~(\ref{eq:AC_Schrodinger_eq}) with an  
infinitesimal variation of the higher-seniority strength parameter
$\lambda$, the first-order
derivative of the integrand can be expressed analytically as follows, 
\be\label{eq:Wc_deriv}
\dfrac{\partial
\overline{\mathcal{W}}^{\mathcal{S},\lambda}}{\partial \lambda}&=&
2 \mel{\Psi_0^\lambda}{\hat{\mathcal{V}}}{\frac{\partial
\Psi_0^\lambda}{ \partial \lambda}}
\\
\label{eq:Wc_deriv_PT1_wf_expansion}
&=&
2 \sum_{I>0}
\dfrac{\bra{\Psi_0^\lambda}\hat{\mathcal{V}}\ket{\Psi_I^\lambda}
\bra{\Psi_I^\lambda}\frac{\partial \hat{H}^\lambda}{\partial
\lambda}\ket{\Psi_0^\lambda}}{\mathcal{E}_0^\lambda -
\mathcal{E}_I^\lambda},
\ee
or, more explicitly [see Eqs.~(\ref{eq:Hamilt_along_AC}) and (\ref{eq:Hlambda_S0})],
\be
\begin{split}
\dfrac{\partial
\overline{\mathcal{W}}^{\mathcal{S},\lambda}}{\partial \lambda}
=2 \sum_{I>0}
&
\dfrac{\bra{\Psi_0^\lambda}\hat{\mathcal{V}}\ket{\Psi_I^\lambda}}
{\mathcal{E}_0^\lambda - \mathcal{E}_I^\lambda}
\\
&\times
\mel{\Psi_I^\lambda}{\hat{\mathcal{V}}+\sum_p \frac{\partial
\varepsilon_p^\lambda}{\partial   \lambda} \hat{n}_p}{\Psi_0^\lambda}
,
\end{split}
\ee
where the summation runs over the excited states of $\hat{H}^\lambda$.\\
Note that, in the $\lambda=0$ limit, the excited states that contribute to the
perturbation expansion should be coupled to the seniority-zero ground-state wave function $\Psi^{\mathcal{S}_0}$ through the higher-seniority
Hamiltonian term $\hat{\mathcal{V}}$:
\be
\mel{\Psi^{\mathcal{S}_0}}{\hat{\mathcal{V}}}{\Psi_I^{\lambda=0}}\neq 0.
\ee  
Obviously, pure seniority-zero solutions will be automatically excluded
from the expansion. In other words, only pure higher-seniority
solutions~\cite{PRB65_Chasman_seniority-zero_excited-states,PR66_Richardson_seniority_zero_solutions,PR67_Richardson_Seniority-one_and_two_solutions} to
$\hat{H}^{\lambda=0}$ will contribute. Note also that, in complete
analogy with G\"{o}rling--Levy PT~\cite{gorling1993correlation,gorling1994exact}, the integrand derivative can be
rewritten as a PT2 energy correction. 
Indeed, according to the natural orbital occupations constraint of
Eq.~(\ref{eq:AC_occ_constraint}),
\be
0=\dfrac{\partial \expval{\hat{n}_p}_{\Psi^\lambda_0}}{\partial
\lambda}=2\mel{\Psi_0^\lambda}{\hat{n}_p}{\frac{\partial
\Psi_0^\lambda}{ \partial \lambda}},
\ee
thus leading to
\be
\begin{split}
\mel{\Psi_0^\lambda}{\hat{\mathcal{V}}}{\frac{\partial
\Psi_0^\lambda}{ \partial \lambda}}&=
\mel{\Psi_0^\lambda}
{
\hat{\mathcal{V}}+\sum_p \frac{\partial
\varepsilon_p^\lambda}{\partial   \lambda} \hat{n}_p
}{\frac{\partial
\Psi_0^\lambda}{ \partial \lambda}
}
\\
&=
\mel{\Psi_0^\lambda}{\dfrac{\partial \hat{H}^\lambda}{\partial
\lambda}}{\frac{\partial
\Psi_0^\lambda}{ \partial \lambda}},
\end{split}
\ee
and, consequently [see Eq.~(\ref{eq:Wc_deriv})], 
\be\label{eq:first_deriv_integrand_derivHamil}
\dfrac{\partial\overline{\mathcal{W}}^{\mathcal{S},\lambda}}{\partial \lambda}=2\mel{\Psi_0^\lambda}{\dfrac{\partial \hat{H}^\lambda}{\partial
\lambda}}{\frac{\partial
\Psi_0^\lambda}{ \partial \lambda}}.
\ee
Finally, from the first-order perturbation expansion in $\lambda$ of $\Psi_0^\lambda$ [see
Eq.~(\ref{eq:Wc_deriv_PT1_wf_expansion})], we do recover (with a factor
2) a second-order energy correction: 
\be
\label{eq:deriv_AC}
\begin{split}
\dfrac{\partial
\overline{\mathcal{W}}^{\mathcal{S},\lambda}}{\partial \lambda}
&=
2 \sum_{I>0}
\dfrac{\left\vert\mel{\Psi_0^\lambda}{
\frac{\partial \hat{H}^\lambda}{\partial
\lambda}
}{\Psi_I^\lambda}\right\vert^2}
{\mathcal{E}_0^\lambda - \mathcal{E}_I^\lambda}
\\
&=
2 \sum_{I>0}
\dfrac{\left\vert\mel{\Psi_0^\lambda}{\hat{\mathcal{V}}
+\sum_p \frac{\partial
\varepsilon_p^\lambda}{\partial   \lambda} \hat{n}_p
}{\Psi_I^\lambda}\right\vert^2}
{\mathcal{E}_0^\lambda - \mathcal{E}_I^\lambda}
.
\end{split}
\ee
If we return to Eq.~(\ref{eq:integrand_PT2}), integrating over the range
$0\leq \lambda\leq 1$ [see Eq.~(\ref{eq:exact_GS_ener_trueAC})] leads to
the following PT2-like total energy expression,
\be\label{eq:PT2_total_ener}
E\overset{\rm
PT2}{\approx}\expval{\hat{H}}_{\Psi^{\mathcal{S}_0}}+\dfrac{1}{2}\left.\dfrac{\partial
\overline{\mathcal{W}}^{\mathcal{S},\lambda}}{\partial \lambda}
\right|_{\lambda = 0}.
\ee 

As an alternative to PT2 for evaluating higher-seniority correlation
energies we may use linear interpolations (LIs) of the AC integrand. In the simplest LI, we 
interpolate between the seniority-zero ($\lambda=0$) and
fully-interacting ($\lambda=1$) limits. We refer to this approximation as
{\it one-segment} LI (1LI):  
\begin{eqnarray}\label{eq:integrand_1LI}
\overline{\mathcal{W}}^{\mathcal{S},\lambda}\overset{\rm 1LI}{\approx}
\lambda\, \overline{\mathcal{W}}^{\mathcal{S},\lambda = 1},
\end{eqnarray}
which gives the following energy expression after integration over
$\lambda$,
\be\label{eq:compfun_1LI_total_ener}
E\overset{\rm 1LI}{\approx}\expval{\hat{H}}_{\Psi^{\mathcal{S}_0}}+\dfrac{1}{2}\,\overline{\mathcal{W}}^{\mathcal{S},\lambda
= 1}.
\ee
We also consider in the present work a refined {\it two-segment} LI
(2LI) where we first linearly interpolate the integrand in the range $0\leq \lambda\leq
1/2$ and then in $1/2\leq \lambda\leq 1$, \ie, 
\begin{eqnarray}\label{eq:integrand_2LI}
\begin{split}
\overline{\mathcal{W}}^{\mathcal{S},\lambda}
&\overset{\rm
2LI}{\approx}
2\lambda \overline{\mathcal{W}}^{\mathcal{S},\lambda =
1/2}\,\mathcal{I}_{[0,1/2[}(\lambda)
\\
&
\quad +\Bigg[2\lambda \left(\overline{\mathcal{W}}^{\mathcal{S},\lambda = 1} - \overline{\mathcal{W}}^{\mathcal{S},\lambda = 1/2}\right) 
\\
&\quad\quad
+ 2 \overline{\mathcal{W}}^{\mathcal{S},\lambda = 1/2}  -  \overline{\mathcal{W}}^{\mathcal{S},\lambda = 1}
\Bigg]\times\mathcal{I}_{[1/2,1]}(\lambda)
,
\end{split}
\end{eqnarray}
where the indicator function is defined as follows,
\begin{eqnarray}\label{step_function}
\mathcal{I}_{A}(\lambda)=\left\{
  \begin{array}{l l}
  1   & \quad {\lambda}\in A \\
  0   & \quad {\lambda}\notin A\\
            \end{array} \right.
.
\end{eqnarray}
The corresponding 2LI energy, which is obtained by integrating over
$0\leq\lambda\leq1$, reads
\be\label{eq:compfun_2LI_total_ener}
E\overset{\rm 2LI}{\approx}\expval{\hat{H}}_{\Psi^{\mathcal{S}_0}}+
\dfrac{1}{2}\overline{\mathcal{W}}^{\mathcal{S},\lambda = 1/2}
+ \dfrac{1}{4}\overline{\mathcal{W}}^{\mathcal{S},\lambda = 1}.
\ee
The performance of the various approximations we have introduced within
the present constrained AC (\ie, 1LI, 2LI, and PT2) will be discussed in detail in Sec.~\ref{sec:discussion}.

\subsection{Relaxed adiabatic connection}
\label{sec:relaxedAC}

For analysis purposes, we consider another (simpler) AC formalism where
the $\lambda$-independent physical potential
$\varepsilon^{\lambda=1}_p=h_{pp}$ is employed along the AC path, thus
relaxing the
natural orbital occupations constraint of
Eq.~(\ref{eq:AC_occ_constraint}). In this {\it relaxed} AC, the
partially-interacting Schr\"{o}dinger equation reads
\be\label{eq:relaxedAC_hamilt}
\left(\sum_p h_{pp} \hat{n}_p + \hat{W}_{\rm
ee}^{\mathcal{S}_0}+\lambda
\hat{\mathcal{V}}\right)\ket{\tilde{\Psi}_0^\lambda}=\tilde{\mathcal{E}}_0^\lambda\ket{\tilde{\Psi}_0^\lambda}.
\ee
The $\lambda=0$ case corresponds to a DOCI-like calculation, where a
straight minimization of the energy $\expval{\hat{H}}$ is performed over seniority-zero
wave functions (constructed here from the exact natural orbitals):
\be\label{eq:relaxedAC_lambda0_ener}
\tilde{\mathcal{E}}_0^{\lambda=0}=\min_{\Psi\in\mathcal{S}_0}\expval{\hat{H}}_{\Psi}=\expval{\hat{H}}_{\tilde{\Psi}^{\mathcal{S}_0}}.
\ee
Unlike the reference seniority-zero wave function
${\Psi}^{\mathcal{S}_0}$ introduced previously, which is a functional of the ground-state
density matrix, the minimizing $\tilde{\Psi}^{\mathcal{S}_0}
=\tilde{\Psi}_0^{\lambda=0}$ solution does {\it not} necessarily
reproduce the exact natural orbital occupancies. Since the true
(physical) Schr\"{o}dinger equation is recovered from
Eq.~(\ref{eq:relaxedAC_hamilt}) when $\lambda=1$, the {\it exact} ground-state energy can be expressed as
follows, 
\be
E&=&\tilde{\mathcal{E}}_0^{\lambda=1}=\tilde{\mathcal{E}}_0^{\lambda=0}+\displaystyle\int^1_0
d\lambda\,\dfrac{\partial
\tilde{\mathcal{E}}_0^\lambda}{\partial\lambda},
\ee
or, equivalently [see Eq.~(\ref{eq:relaxedAC_lambda0_ener})],
\be
E=\expval{\hat{H}}_{\tilde{\Psi}^{\mathcal{S}_0}}+\displaystyle\int^1_0
d\lambda\,\tilde{\overline{\mathcal{W}}}^{\mathcal{S},\lambda},
\ee
where, according to the Hellmann--Feynman theorem and
Eq.~(\ref{eq:relaxedAC_hamilt}), the relaxed AC integrand reads
\be
\tilde{\overline{\mathcal{W}}}^{\mathcal{S},\lambda}=\dfrac{\partial
\tilde{\mathcal{E}}_0^\lambda}{\partial\lambda}=\expval{\hat{\mathcal{V}}}_{\tilde{\Psi}_0^{\lambda}}.
\ee
By analogy with Eqs.~(\ref{eq:integrand_PT2}) and
(\ref{eq:PT2_total_ener}), an approximate PT2 scheme can be derived
along the relaxed AC if we assume that the integrand varies linearly in
$\lambda$,
\ie, 
\begin{eqnarray}\label{eq:integrand_relaxedPT2}
\tilde{\overline{\mathcal{W}}}^{\mathcal{S},\lambda }
&\overset{\rm PT2}{\underset{\mbox{\footnotesize relaxed AC}}{\approx}}&
 \lambda \left.\dfrac{\partial
\tilde{\overline{\mathcal{W}}}^{\mathcal{S},\lambda}}{\partial \lambda}
\right|_{\lambda = 0},
\end{eqnarray}
thus leading to the following PT2-like energy expression after integration over
$\lambda$:
\be\label{eq:PT2_total_ener_relaxed}
E
&\overset{\rm PT2}{\underset{\mbox{\footnotesize relaxed AC}}{\approx}}&
\expval{\hat{H}}_{\tilde{\Psi}^{\mathcal{S}_0}}
+\dfrac{1}{2}\left.\dfrac{\partial
\tilde{\overline{\mathcal{W}}}^{\mathcal{S},\lambda}}{\partial \lambda}
\right|_{\lambda = 0}.
\ee
According to perturbation theory that we apply to the relaxed AC Hamiltonian of 
Eq.~(\ref{eq:relaxedAC_hamilt}) with an infinitesimal variation of
$\lambda$, the first-order derivative of the relaxed AC integrand reads
more explicitly 
\be
\label{eq:ACint_relax_derive}
\dfrac{\partial
\tilde{\overline{\mathcal{W}}}^{\mathcal{S},\lambda}}{\partial \lambda}=
2 \sum_{I>0}
\dfrac{\bra{\tilde{\Psi}_0^\lambda}\hat{\mathcal{V}}\ket{\tilde{\Psi}_I^\lambda}
\bra{\tilde{\Psi}_I^\lambda}\hat{\mathcal{V}}\ket{\tilde{\Psi}_0^\lambda}}{\tilde{\mathcal{E}}_0^\lambda
- \tilde{\mathcal{E}}_I^\lambda}.
\label{eq:deriv_relaxedAC}
\ee
Note that, unlike the 1LI or 2LI approximations that have been
introduced in Sec.~\ref{subsec:approx_AC}, PT2 relies solely on pure zero-
and higher-seniority wave functions. As further discussed in Sec.~\ref{sec:discussion}, in relatively weak 
correlation regimes, PT2 systematically 
overestimates the higher-seniority correlation
energy, when applied to the
constrained AC. Its performance dramatically improves when the relaxed AC
path is followed instead.  


\section{Computational details}\label{sec:comput_details}

The reverse engineering procedure described in
Eq.~(\ref{eq:AC_Lieb_max}), where the true natural orbital occupancies
$\left\{n_p^{\Psi_0}\right\}$
are used as input in the Lieb maximization, has been
implemented
together with the Block {\it density matrix renormalization group}
(DMRG)
code~\cite{chan2002highly,chan2004algorithm,ghosh2008orbital,sharma2012spin,olivares2015ab,guo2016n}.
The DMRG code is used to obtain the ground-state energy
of the
$\lambda$-dependent Hamiltonian
[see Eqs.~(\ref{eq:AC_Schrodinger_eq}) and (\ref{eq:Hlambda_S0})]
for any $\lambda$ in $0\leq \lambda\leq 1$.
The natural orbitals were obtained by diagonalizing the 1RDM from the reference calculation (FCI or DMRG) in the canonical orbital basis.
The DIIS algorithm has been used to find a set of potential values $\lbrace
\varepsilon^\lambda_p \rbrace$ which satisfies
\be\label{eq:LFT_opt}
n_p^{\Psi_0^\lambda\left(\left\{\varepsilon^\lambda_p\right\}\right)}-n_p^{\Psi_0}\overset{!}{=}0.
\ee
The threshold for convergence of the DMRG sweep energies has been set
to $10^{-10}$ Hartree, and threshold for the Euclidian norm of the residual in Eq.~(\ref{eq:LFT_opt}) has been set to $10^{-5}$.
As a proof of concept, we test our method on the metallic H$_4$ linear chain (considering symmetric stretching) in
the STO-3G basis, the metallic H$_8$ linear chain in the cc-pVDZ basis,
and the Helium dimer~\cite{mentel2012formulation} in the cc-pVDZ basis.
In the latter case, no basis set superposition error (BSSE) corrections have
been applied so that a direct comparison can be made with the results of
Ref.~\cite{zoboki2013linearized}.
The number of renormalized states in the DMRG calculation is
set to $m=400$ for He$_2$, and $m=800$ for H$_8$.
In the latter case, the DMRG energy with respect to $m$ converged with an order of $10^{-5}$ Hartree (see Appendix~\ref{app:convergence_DMRG}).
Molcas~\cite{aquilante2016molcas} has been used to compute the one- and two-body electronic integrals.
In order to further analyze the AC integrand in the case of the H$_4$
chain in the minimal basis (in particular its variation along the AC), Psi4~\cite{psi4} was used to generate the electronic integrals and OpenFermion~\cite{openfermion} was used to construct the ($\lambda$-dependent) Hamiltonian that is then diagonalized exactly in the subspace conserving the number of particles and spin singlet.

\section{Results and discussion}\label{sec:discussion}

\subsection{H$_4$ chain in a minimal basis}

\subsubsection{AC and potential energy curves}

\begin{figure}
\centering
\resizebox{\columnwidth}{!}{
\includegraphics[scale=1]{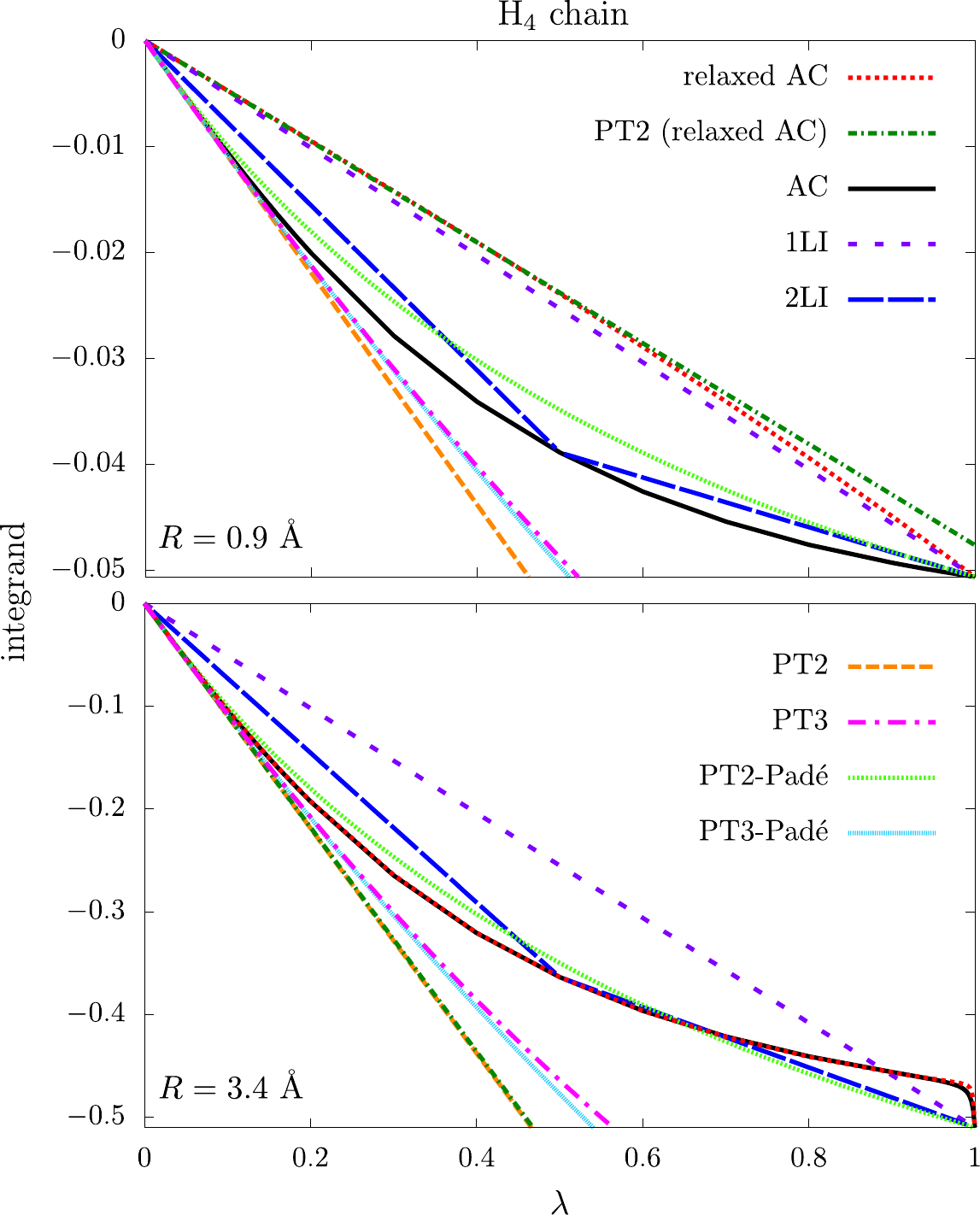}
}
\caption{``Exact'' and approximate higher-seniority correlation
integrands evaluated along the constrained (simply referred to as AC) and relaxed ACs for the
H$_4$ chain with interatomic distances $R=0.9$~{\AA} (top panel) and
$R=3.4$~{\AA} (bottom panel). Energies are in atomic unit (hartree).
See text for further details.}
\label{fig:Wc_H4chain}
\end{figure}

Let us start with the H$_4$ linear chain in the minimal basis STO-3G.
The \exact ({\it i.e.}, exact in the considered minimal basis) and approximate integrands are shown in Fig.~\ref{fig:Wc_H4chain}
for interatomic distances $R = 0.9$~{\AA} (top panel) and
$R = 3.4$~{\AA} (bottom panel) corresponding to a weakly and strongly
correlated regimes, respectively.
In analogy with the one obtained in the approach of Pernal~\cite{pernal2018exact}, our exact AC integrand has a quadratic behaviour with respect to $\lambda$.
However, it shows a much more pronounced curvature so that the 1LI
approximation, the PT2 approximation, or the linear extrapolation
schemes proposed in Ref.~\cite{pernal2018electron}, are expected to be
less accurate.
To improve over the PT2 approximation, the PT3 approximation as well as Pad\'{e} approximants have been considered for the constrained AC, and are detailed in Appendix~\ref{app:pade}.
As readily seen in Fig.~\ref{fig:Wc_H4chain}, while the PT3 and PT3-Padé approximations show some curvature, they are still not sufficiently accurate.
In contrast, the PT2-Padé approximant -- which is exact in the $\lambda \rightarrow 0$ and
$\lambda \rightarrow 1$ limits of the AC integrand -- recovers a large part of the curvature by construction.  \\

\begin{figure}
\centering
\resizebox{\columnwidth}{!}{
\includegraphics[scale=1]{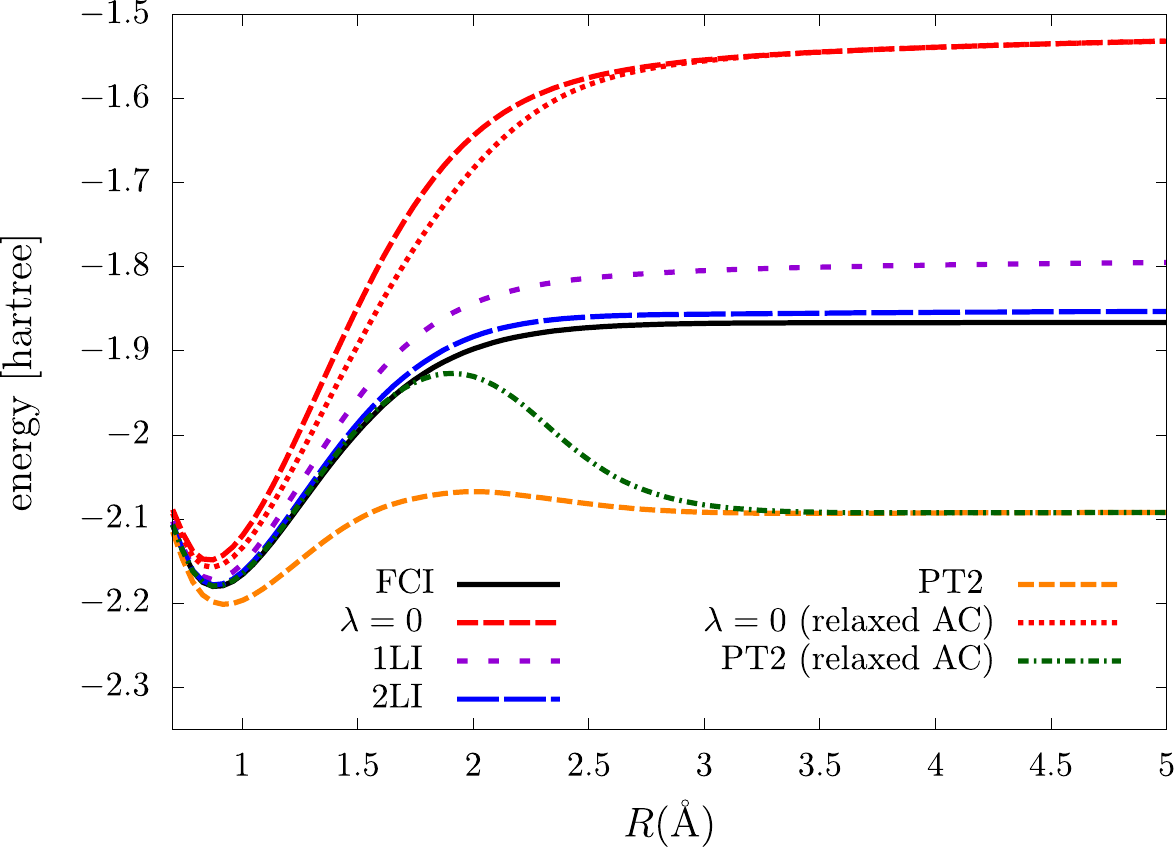}
}
\caption{Potential energy curves plotted for the H$_4$ chain (with
equidistant hydrogen atoms) at both
seniority-zero ($\lambda=0$), which is equivalent to perturbation theory
through first order, and higher-seniority (1LI, 2LI, or PT2) levels of
approximation within the constrained AC. Results obtained from the
relaxed AC are shown for analysis purposes. See text for further
details.}
\label{fig:diss_H4}
\end{figure}

We show in Fig.~\ref{fig:diss_H4} the potential energy curves generated
from both the AC with the natural orbital occupation constraint (that we
simply refer to as constrained AC in the following) and its relaxed
version, with and without higher-seniority correlation energy
corrections [see Eqs.~(\ref{eq:PT2_total_ener}), (\ref{eq:compfun_1LI_total_ener}),
(\ref{eq:compfun_2LI_total_ener}), and (\ref{eq:PT2_total_ener_relaxed})].
Equilibrium bond distances and total energies as well as dissociation energies
are also provided in Table~\ref{tab:diss_H4}. 
We see that
the relaxed and
constrained seniority-zero ($\lambda=0$) wave functions slightly
underestimate the bond distance (in comparison with FCI). However, they both
overestimate (by about 100\%) the dissociation energy. 
The PT2 approximation to the constrained AC systematically
overestimates (roughly by 100\% for all bond distances) the higher-seniority correlation energy, with a large
error on the total energy when the chain is stretched. 
As shown in
Table~\ref{tab:diss_H4}, at this level of approximation, the equilibrium bond
distance is too long (in comparison with FCI) and the dissociation energy
is dramatically underestimated.    
On the other hand, approximating the 
constrained AC integrand with a linear interpolation over the entire
interaction strength $0\leq \lambda\leq 1$ (1LI approximation) gives
much better (although too high) total energies. 
The equilibrium bond distance is relatively
accurate in this case (see Table~\ref{tab:diss_H4}). The error
increases along the dissociation coordinate $R$ but remains substantially
smaller than that of PT2. The 1LI approximation substantially improves the
description of the dissociation energy. 
Linearly interpolating over the two segments
$0\leq \lambda \leq 1/2$ and $1/2\leq \lambda\leq  1$ (2LI
approximation) is, by construction, designed to better account for the
quadratic behaviour of the constrained AC integrand (see
Fig.~\ref{fig:Wc_H4chain}), thus providing more accurate total energies
for all bond distances, as expected.\\

Let us now focus on the relaxed AC results. Unlike in the constrained
AC, the DOCI coefficients of the reference seniority-zero wave function
are determined by energy minimization. They are not expected to
reproduce the \exact natural orbital occupancies but will obviously give
a lower (although not substantially better, even at equilibrium) total energy when the higher-seniority correlation energy is
neglected (see the ``$\lambda=0$'' curves in Fig.~\ref{fig:diss_H4}).
More interestingly, around equilibrium, the relaxed AC integrand exhibits a
very weak curvature (see the top panel of Fig.~\ref{fig:Wc_H4chain}) 
Therefore, the PT2 approximation to the relaxed AC (see Fig.~\ref{fig:diss_H4}) gives very
accurate total energies for bond distances shorter than $R=1.75$~\AA. Beyond
this distance the results dramatically deteriorate and, for $R >
3.4$~{\AA}, the relaxed and constrained ACs give the same total energies
at the PT2 level of approximation. As a result, the PT2 description of
the dissociation
energy worsens even further when relaxing the AC (see Table~\ref{tab:diss_H4}). As shown in Fig.~\ref{fig:diss_H4}, in the latter case, the
seniority-zero total energy is the same (whether the AC is relaxed or
not). Moreover, for $R=
3.4$~{\AA}, the two AC integrands are indistinguishable not only around
$\lambda=0$, as expected from the PT2 results, but also for larger
higher-seniority interaction strength values $\lambda$ (see the bottom panel of
Fig.~\ref{fig:Wc_H4chain}).
Slight differences only appear in the vicinity of 
$\lambda=1$.

\subsubsection{Potential and integrand slope along the AC}

\begin{figure}
\centering
\resizebox{\columnwidth}{!}{
\includegraphics[scale=1]{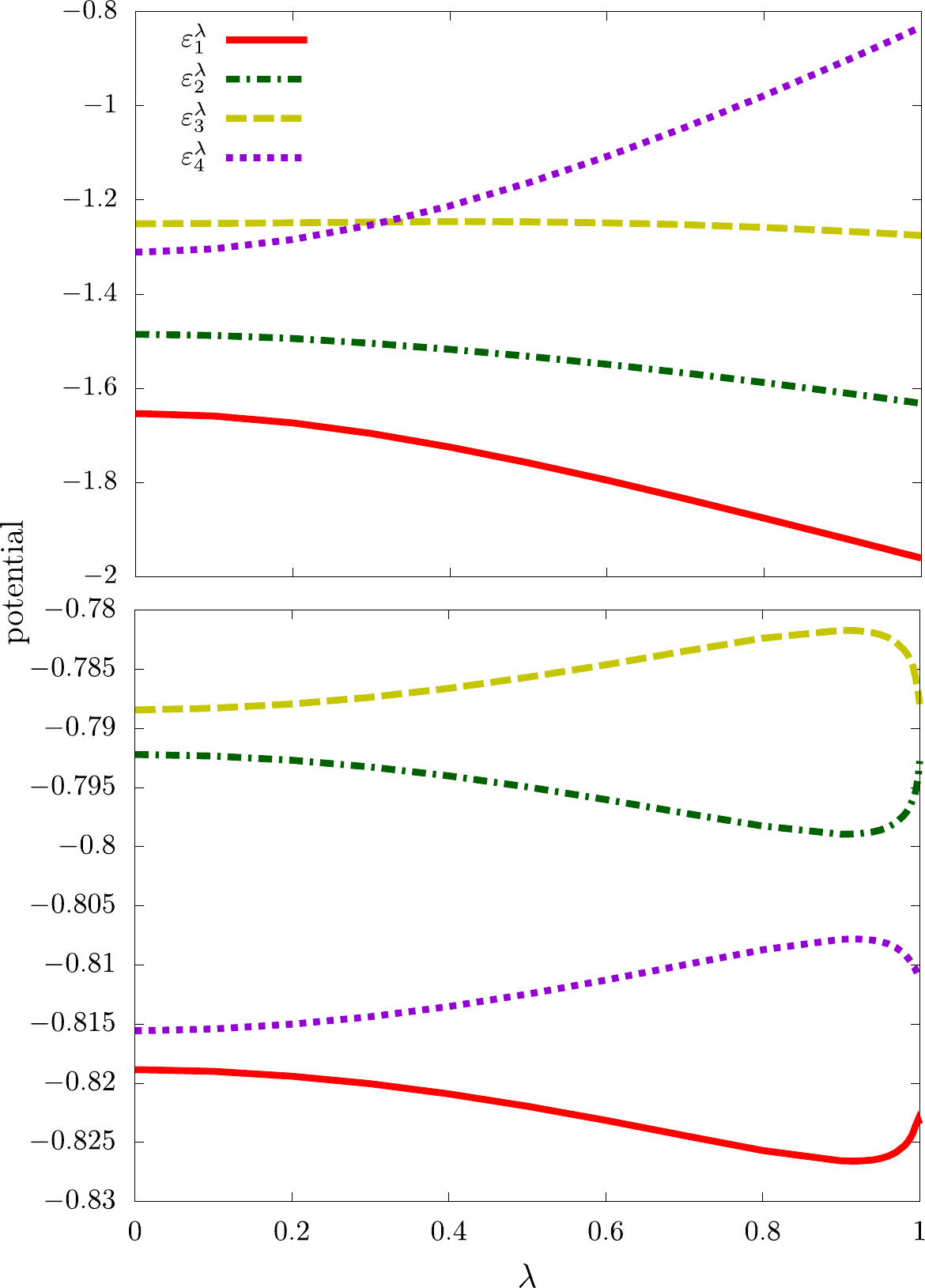}
}
\caption{Potential values $\left\{\varepsilon^\lambda_p\right\}$
computed along
the constrained AC [see Eq.~(\ref{eq:AC_Lieb_max})] for the H$_4$ molecular chain in a minimal basis with
interatomic distances $R=0.9$~{\AA} (top panel) and $R=3.4$~{\AA}
(bottom panel). Energies are in atomic unit (hartree).}
\label{fig:potential_H4chain}
\end{figure}

For further comparison of the constrained and relaxed ACs
at equilibrium and stretched geometries, we show in
Fig.~\ref{fig:potential_H4chain} \exact potential values $\varepsilon_p^\lambda$ over the
natural orbital space ($1\leq p\leq 4$) and along the constrained AC
(\ie, for $0\leq \lambda\leq 1$). As readily seen from the
top panel of Fig.~\ref{fig:potential_H4chain}, variations in $\lambda$
are quite substantial when $R=0.9$~{\AA}. We recall that the
$\lambda$-independent potential
$\left\{\varepsilon_p^{\lambda=1}\right\}$ is employed in the
relaxed AC, thus suggesting that constrained and relaxed AC curves 
should indeed be different at equilibrium (the impact of the
potential on the shape of the AC integrand will be further rationalized
in the following). In contrast, in the stretched
$R=3.4$~{\AA} geometry, the dependence in $\lambda$ of the potential is
drastically reduced (see the bottom panel of
Fig.~\ref{fig:potential_H4chain}). Indeed, in this case, the deviation
$\abs{\varepsilon_p^\lambda-\varepsilon_p^{\lambda=1}}$
from the fully-interacting ($\lambda=1$) potential never exceeds 7
millihartrees, against 0.5 hartrees at equilibrium (see the top panel of
Fig.~\ref{fig:potential_H4chain}). This explains why the constrained and
relaxed AC curves are essentially on top of each other when
$R=3.4$~{\AA} (see the bottom
panel of Fig.~\ref{fig:Wc_H4chain}). At this point we should emphasize
that the natural orbitals become singly occupied in the dissociation
limit. As a result, in the present minimal basis, the density matrix
becomes the identity matrix and the natural orbitals can
be arbitrarily rotated. In the streched $R=3.4$~{\AA} geometry, the
natural orbital occupations are closed (but not strictly equal) to 1.
Even though the natural orbitals are slightly more localized, their
expansion over the atomic orbitals ressembles that of the canonical HF
orbitals in this case. 

\begin{figure}
\centering
\resizebox{\columnwidth}{!}{
\includegraphics[scale=1]{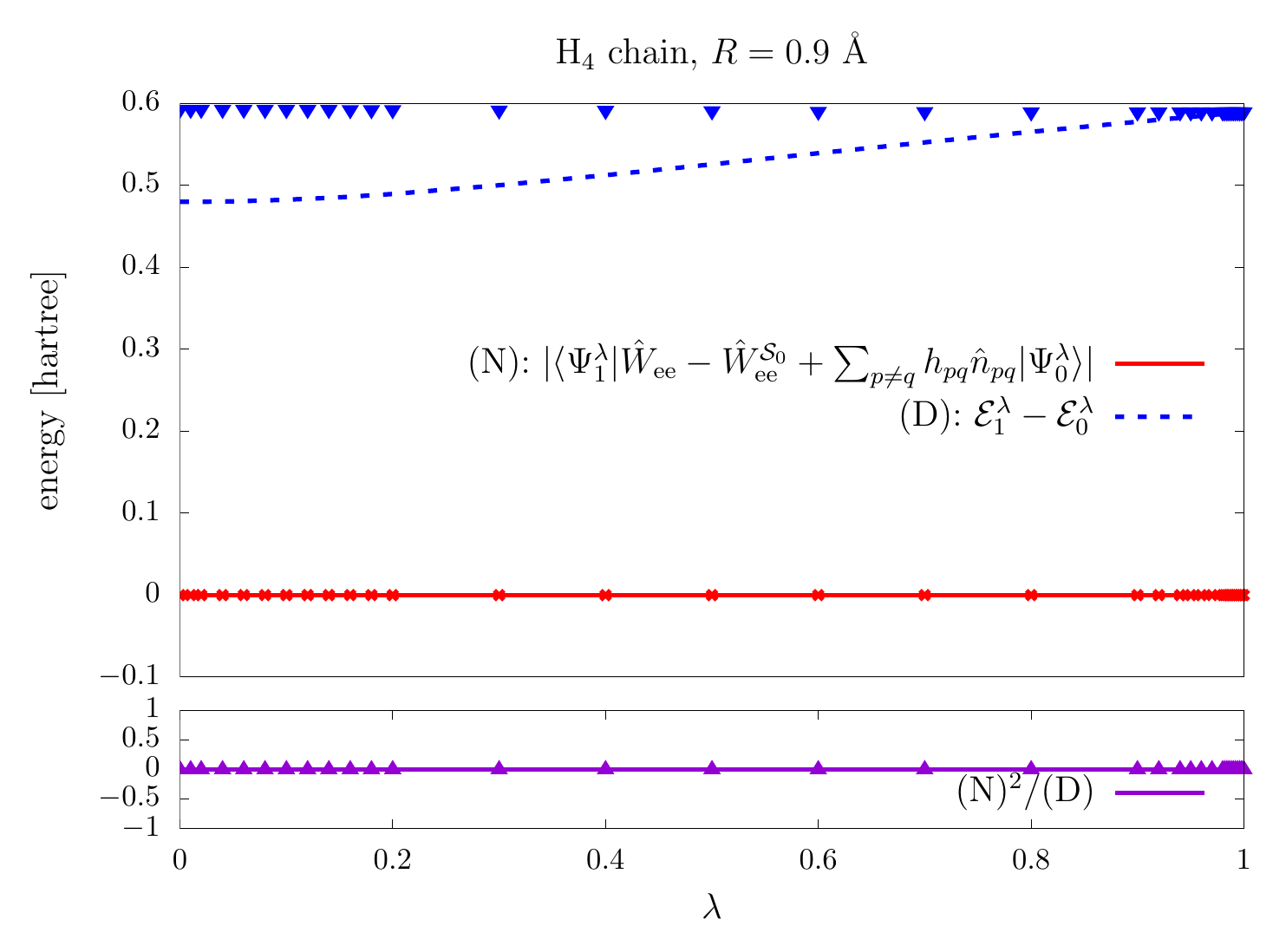}
}
\resizebox{\columnwidth}{!}{
\includegraphics[scale=1]{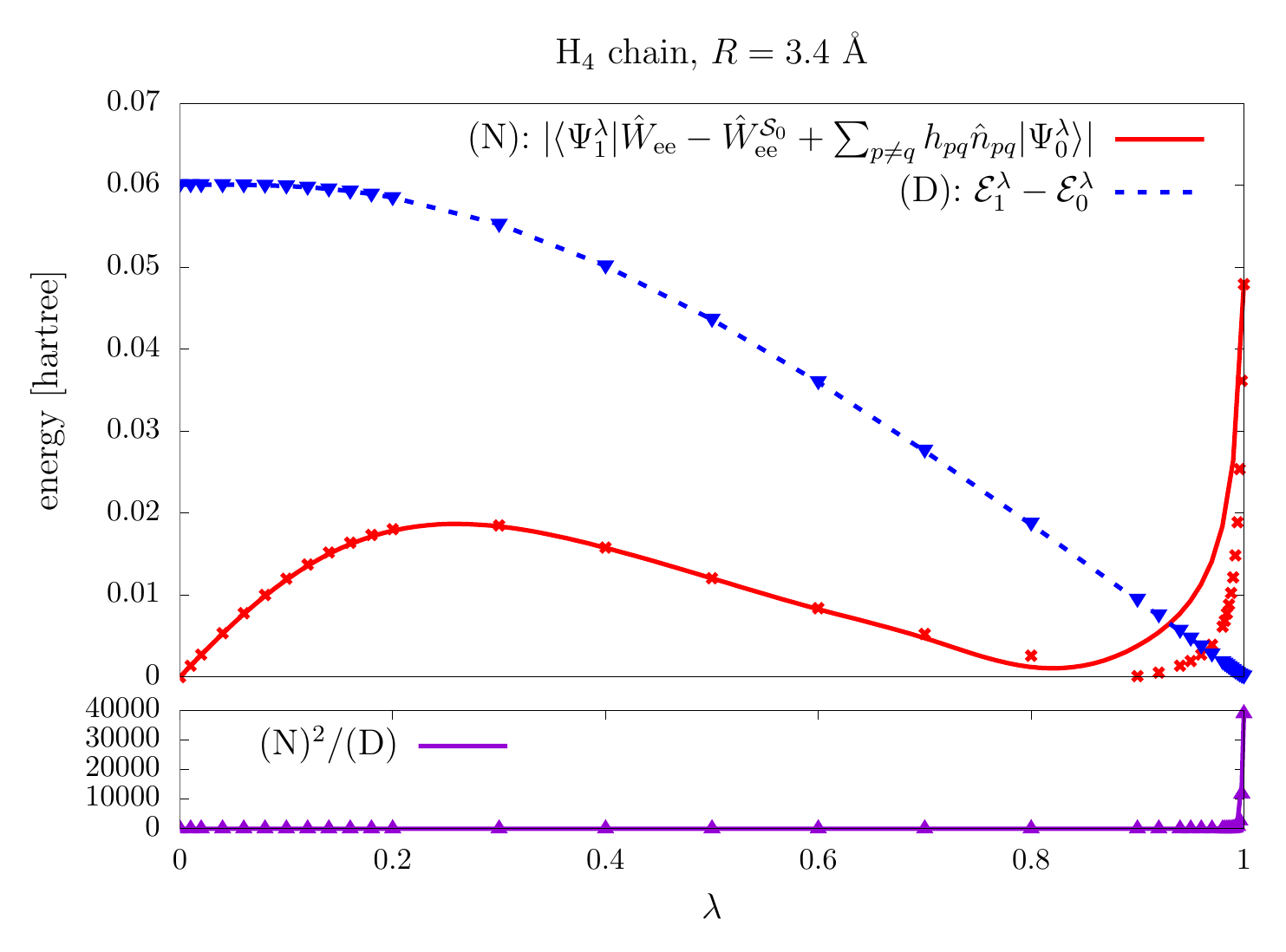}
}
\caption{Detailed first-excited-state contribution to
the derivative in $\lambda$ of the constrained (solid and dashed lines) and relaxed
(solid dots) AC integrands [see Eqs.~(\ref{eq:deriv_AC}) and
(\ref{eq:ACint_relax_derive})]. The two upper and lower panels correspond to
the H$_4$ chain with bond distances $R = 0.9$~{\AA} and $R = 3.4$~{\AA},
respectively. See text for further details.
}
\label{fig:Wc_deriv_10}
\end{figure}

\begin{figure}
\centering
\resizebox{\columnwidth}{!}{
\includegraphics[scale=1]{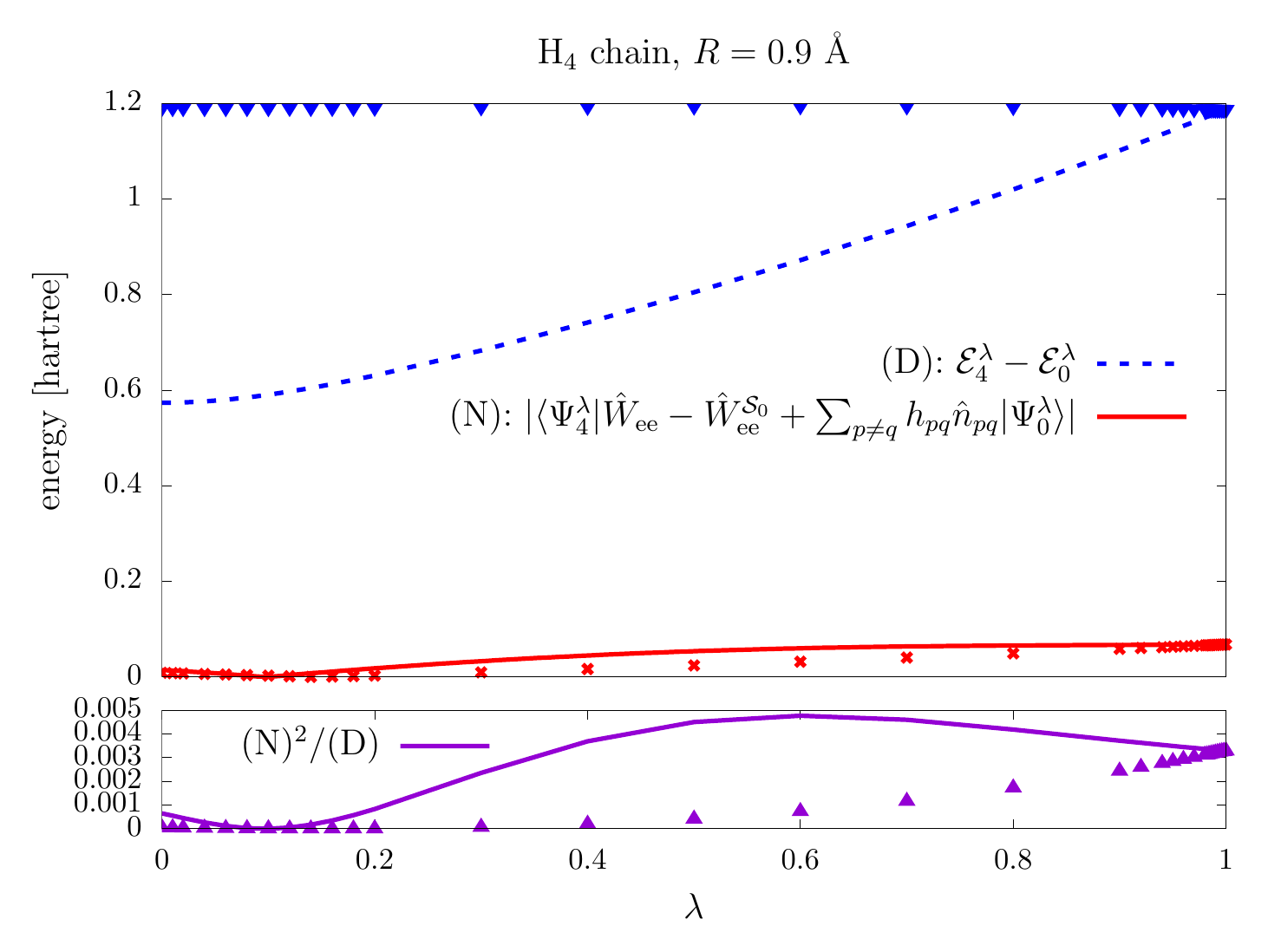}
}
\resizebox{\columnwidth}{!}{
\includegraphics[scale=1]{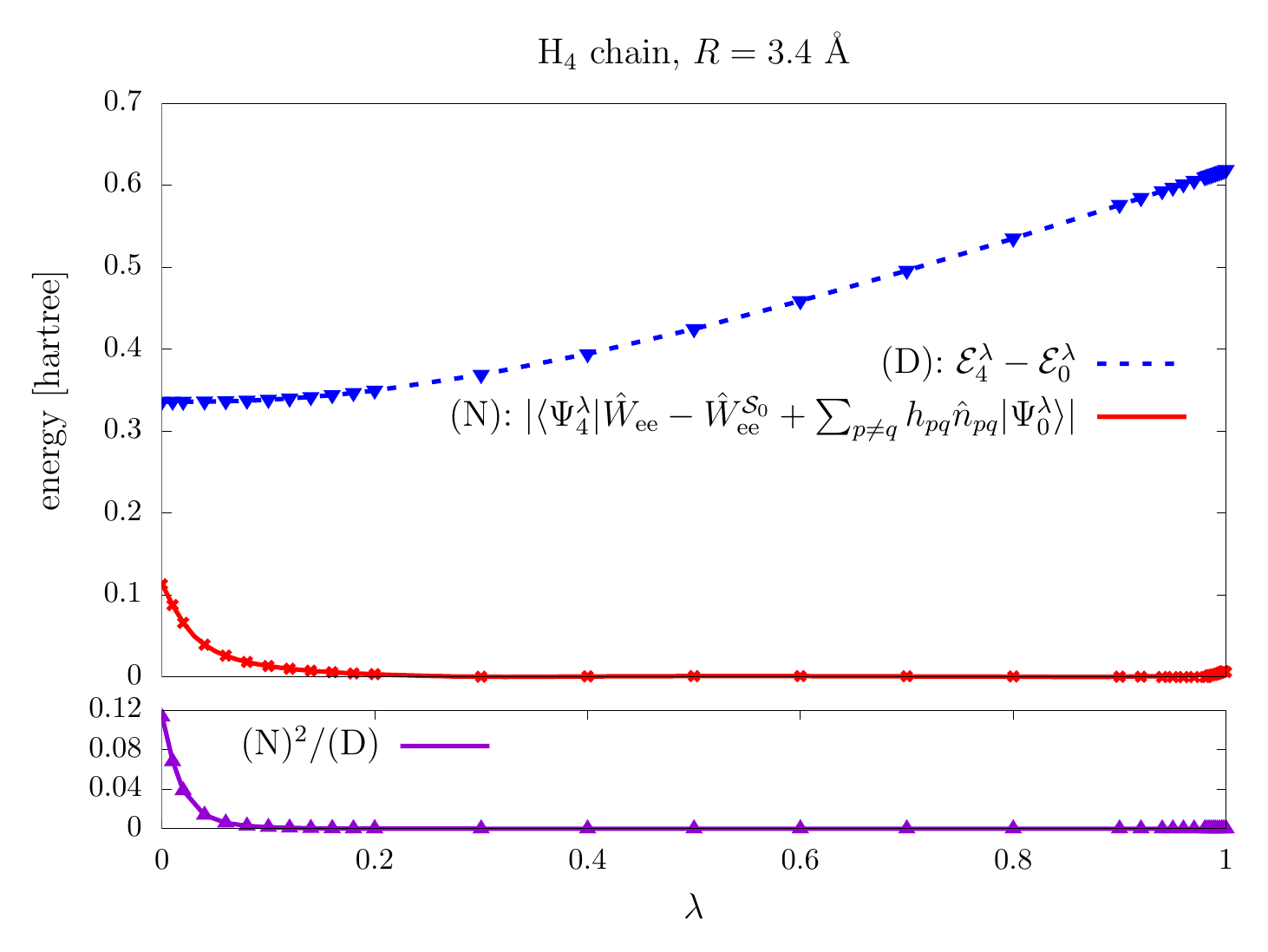}
}
\caption{Same as Fig.~\ref{fig:Wc_deriv_10} for the fourth excited
state. Constrained and relaxed AC results are essentially on top of
each other when $R = 3.4$~{\AA} (two lower panels).}
\label{fig:Wc_deriv_40}
\end{figure}

In order to further rationalize the shape of the constrained and relaxed AC integrands, it is
actually instructive to look at their slopes in $\lambda$ along the AC.
Exact analytical expressions are trivially obtained from static perturbation
theory [see Eqs.~(\ref{eq:deriv_AC}) and (\ref{eq:deriv_relaxedAC}),
respectively]. At equilibrium, the substantial variations in $\lambda$
of the potential along the constrained AC are expected to be reflected
in the excitation energies $\mathcal{E}_I^\lambda -
\mathcal{E}_0^\lambda$ (\ie, the denominators in the perturbation
expansion). We do observe this feature numerically, as shown in the 
top panels of Figs.~\ref{fig:Wc_deriv_10} and \ref{fig:Wc_deriv_40} for
the first and fourth excited state, respectively. Unlike the former, the
latter couples to the ground state along the AC and the resulting
contribution to the perturbation expansion is significant. Note that, as
readily seen in Eqs.~(\ref{eq:deriv_AC}) and (\ref{eq:deriv_relaxedAC}), AC
integrand slopes will always be negative, like a second-order
ground-state energy
correction in perturbation theory. In the
relaxed AC, the fourth excitation energy is much smaller than that of the
constrained AC away from the
fully-interacting $\lambda=1$ limit, especially in the vicinity of the
seniority-zero ($\lambda=0$) limit. This is the reason why the constrained
AC integrand has a much steeper slope at $\lambda=0$ (see the top panel
of Fig.~\ref{fig:Wc_H4chain}). Finally, the expected quasi-independence in
$\lambda$ of the excitation energies along the relaxed AC, which leads
to a relatively weak dependence in $\lambda$ of the integrand slope (especially for $\lambda\leq 0.5$), explains why
the PT2 approximation is relatively accurate in this case (see the top panel
of Fig.~\ref{fig:Wc_H4chain}).   
Turning to the stretched $R=3.4$~{\AA} geometry, the above mentioned
weak dependence in $\lambda$ of the potential along the constrained AC
explains why the relaxed AC essentially gives the same excitation
energies (see the bottom panels of
Figs.~\ref{fig:Wc_deriv_10} and \ref{fig:Wc_deriv_40}) and, ultimately,
to the same AC curves. The slight differences observed in the vicinity
of the fully-interacting 
$\lambda=1$ limit (see the bottom panel of Fig.~\ref{fig:Wc_H4chain})
can be traced back to the coupling (through 
higher-seniority contributions to the Hamiltonian) between the ground and
first excited states, as shown in the bottom panel of
Fig.~\ref{fig:Wc_deriv_10}. We can also relate the sudden drop in slope,
which is a peculiar feature that both constrained and relaxed ACs
exhibit when approaching the latter limit, to both the increase of the
coupling and the drop of the first excitation energy. This feature
appears when the bond distance is larger than $R=2$~{\AA} (not shown). As shown in
Fig.~\ref{fig:diss_H4}, this is precisely the distance at which the  
PT2 description of the constrained AC is not accurate anymore.

\begin{table}
\begin{center}
\begin{tabularx}{\columnwidth}{l*{1}{Y} l*{3}{Y}}
\hline H$_4$
 & \multicolumn{1}{c}{$R_{\rm eq}$ (\AA)}
  & \multicolumn{1}{c}{\vtop{\hbox{\strut Equilibrium}\hbox{\strut energy}}}
    & \multicolumn{1}{c}{\vtop{\hbox{\strut Dissociation}\hbox{\strut energy}}} \\
\hline
$\lambda = 0$ & 0.86 & -2.148979 & 0.617577 \\
$\lambda = 0$ (relaxed AC) & 0.87 &  -2.157561 & 0.626158 \\
PT2 &  0.94 & -2.201520 & 0.109404 \\
PT2 (relaxed AC) &  0.89 & -2.179969 & 0.087854 \\
1LI & 0.88 &  -2.172768 & 0.378017  \\
2LI & 0.89 &  -2.178426 & 0.325506  \\
$\lambda = 1$ (FCI) & 0.89 & -2.180501  & 0.314174  \\
\hline
\end{tabularx}
\caption{Equilibrium bond distance, total energy at equilibrium, and dissociation energy (in Hartree) of the
H$_4$ chain computed in the minimal STO-3G basis at the constrained and
relaxed seniority-zero
($\lambda=0$) levels of
approximation (which is equivalent to perturbation theory through first
order) and beyond. In the latter case, PT2, 1LI or 2LI
energy corrections are applied [see
Eqs.~(\ref{eq:PT2_total_ener}), (\ref{eq:compfun_1LI_total_ener})
and (\ref{eq:compfun_2LI_total_ener}) for the constrained AC, and Eq.~(\ref{eq:PT2_total_ener_relaxed}) for the relaxed AC].
Comparison is made with FCI. The dissociation energy is simply evaluated as the difference between
the energies at $R = 5$~\AA\, and $R=R_{\rm eq}$. See text for further
details.}
\label{tab:diss_H4}
\end{center}
\end{table}


\subsection{H$_8$ chain and Helium dimer}


We now turn to the longer H$_8$ chain and the Helium dimer for which the
constrained AC has been implemented in the cc-pVDZ basis.
As the number of orbitals is too large to perform an exact diagonalization,
we used the DMRG method to perform the Lieb maximization and compute the
corresponding AC integrand.\\

\begin{figure}
\centering
\resizebox{\columnwidth}{!}{
\includegraphics[scale=1]{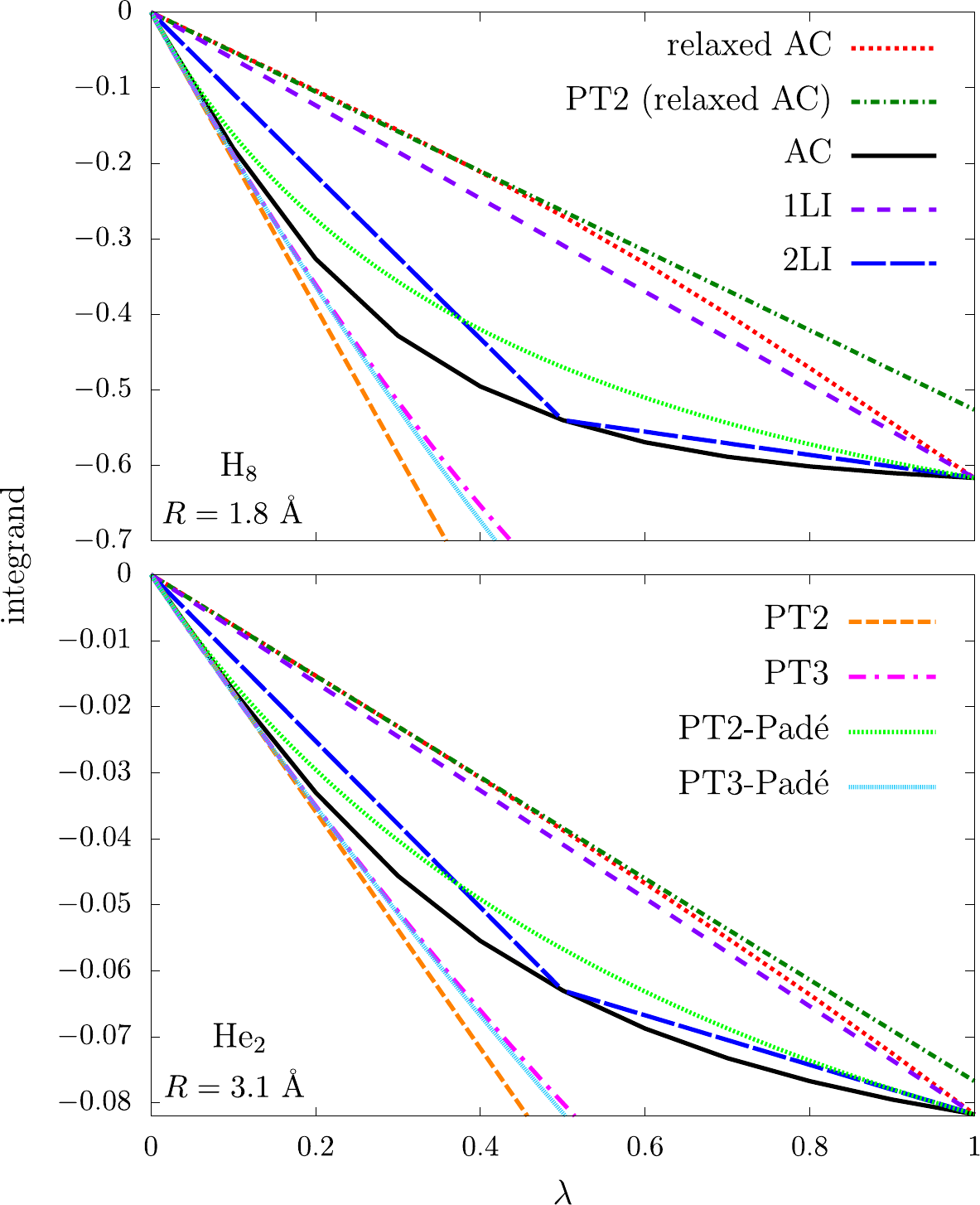}
}
\caption{Constrained and relaxed AC integrands computed at various
levels of approximation for
the H$_8$ molecular chain
(top panel) and
the He$_2$ molecule (bottom panel). DMRG results (simply referred to as ``AC''
and ``relaxed AC'') are
used as reference. Energies are in atomic unit (hartree).
For the large interatomic distance $R = 30$~\AA, the AC integrand of He$_2$ is zero (not shown).}
\label{fig:Wc_allsystems}
\end{figure}

For comparison with Pernal's AC formalism~\cite{pernal2018electron}, we
plot in the top panel of Fig.~\ref{fig:Wc_allsystems} accurate and
approximate AC integrands for
the stretched H$_8$ chain with equidistant hydrogen nuclei ($R=1.8$ \AA).
Many features that have been discussed previously for the H$_4$ chain (at
equilibrium) are essentially recovered. We note, in particular, the
pronounced curvature of the constrained AC integrand that, by
construction, PT2 and 1LI completely miss. As shown in
Table.~\ref{tab:corr}, like in H$_4$ or the stretched He$_2$ molecule
(see also the bottom panel of Fig.~\ref{fig:Wc_allsystems}), the former and latter approximations substantially
overestimate and underestimate the higher-seniority correlation energy,
respectively. As expected, a better description of the constrained AC is
obtained at the 2LI level of approximation (see
Table.~\ref{tab:corr}) or with the PT2-Pad\'{e} approximant (see
Appendix~\ref{app:pade}).\\

\begin{table*}
\begin{center}
\begin{tabularx}{\textwidth}{c *{5}{Y}}
\hline  \multicolumn{1}{c}{$\overline{W}^{\mathcal{S}}$}
  & \multicolumn{1}{c}{H$_8$ chain ($R=1.8$~{\AA})}
    & \multicolumn{1}{c}{He$_2$ ($R=3.1$~{\AA})}
    &  \multicolumn{1}{c}{H$_4$ chain ($R = 0.9$ \AA)}
    & \multicolumn{1}{c}{H$_4$ chain ($R = 3.4$ \AA)}\\
\hline
reference &  -0.466432 & -0.055487 &  -0.034278 &  -0.320665 \\
PT2 &   -0.975657(109\%)&  -0.089689(62\%) & -0.054791(60\%) & -0.547304(71\%)\\
PT3 & -0.702535(51\%) & -0.075878(37\%) & -0.046463(36\%) & -0.433155(35\%) \\
PT2-Pad\'{e} & -0.422700(-9\%) & -0.051760(-7\%) & -0.031894(-7\%) & -0.320182(-0.1\%) \\
PT3-Pad\'{e} & -0.771226(65\%) & -0.078647(42\%) & -0.047984(40\%) & -0.457629(42\%) \\
1LI &  -0.308280(-34\%) &  -0.040870(-26\%) & -0.025332(-26\%) &  -0.255112(-20\%) \\
2LI &  -0.424164(-9\%)& -0.051930(-6\%) & -0.032126(-6\%) & -0.309438(-3\%) \\
\hline  \multicolumn{1}{c}{Total energy}
  & & & & \\
  \hline
Hartree--Fock & -3.817791 & -5.710322 & -2.124260 & -1.268200\\
$\left \langle \hat{H} \right \rangle_{\Psi^{\mathcal{S}_0}}$ & -3.663506 & -5.719709 & -2.146038 & -1.545865 \\
DMRG & -4.129982 & -5.775196 & -- & --\\
FCI & -- & -- & -2.180317 & -1.866530	\\

\end{tabularx}
\caption{Higher-seniority correlation energy
$\overline{W}^{\mathcal{S}}$ computed along the constrained AC at various levels of approximation (see Sec.~\ref{subsec:approx_AC}) for the
H$_8$ chain, the Helium dimer, and the H$_4$ chain. Relative deviations
from the reference results are given in parentheses. The reference
method is DMRG for He$_2$ and H$_8$, and FCI for H$_4$.
The total energy using Hartree--Fock, FCI, and DMRG are also reported, together with the expectation value of the Hamiltonian computed with the converged seniority-zero wavefunction.}
\label{tab:corr}
\end{center}
\end{table*}

If we relax the constraint on the natural
orbital occupations, the curvature of the AC integrand is drastically
reduced, like in the H$_4$ chain at equilibrium, thus leading to a
relatively good description of the (relaxed) AC at both PT2 and 1LI levels
of approximation. Interestingly, similar features have been reported by
Pernal~\cite{pernal2018electron} for the stretched H$_8$
chain within a different AC formalism where, instead, a regular {\it generalized valence
bond} (GVB) wave function is used as reference (see Fig.~2 of
Ref.~\cite{pernal2018electron}). Still, we note that, for $\lambda=1$, the
relaxed AC integrand is (in absolute value) larger by a factor 3 than
that of Pernal. This difference originates from the fact that, in the present
(relaxed or constrained) AC, the reference ($\lambda=0$) seniority-zero
wave function is constructed from the \exact physical natural orbitals (\ie, those obtained in the $\lambda=1$ limit of the
AC). Since orbital optimization is of primary importance in variational 
seniority-zero calculations~\cite{boguslawski2014efficient,limacher2014influence},
relaxing the natural orbitals constraint, like in Pernal's AC~\cite{pernal2018electron}, would further reduce the
higher-seniority correlation energy. The resulting seniority-zero wave
function would be optimal energy wise but it would neither reproduce the
exact natural orbitals nor their occupancies.\\ 

\begin{figure}
\centering
\resizebox{\columnwidth}{!}{
\includegraphics[scale=1]{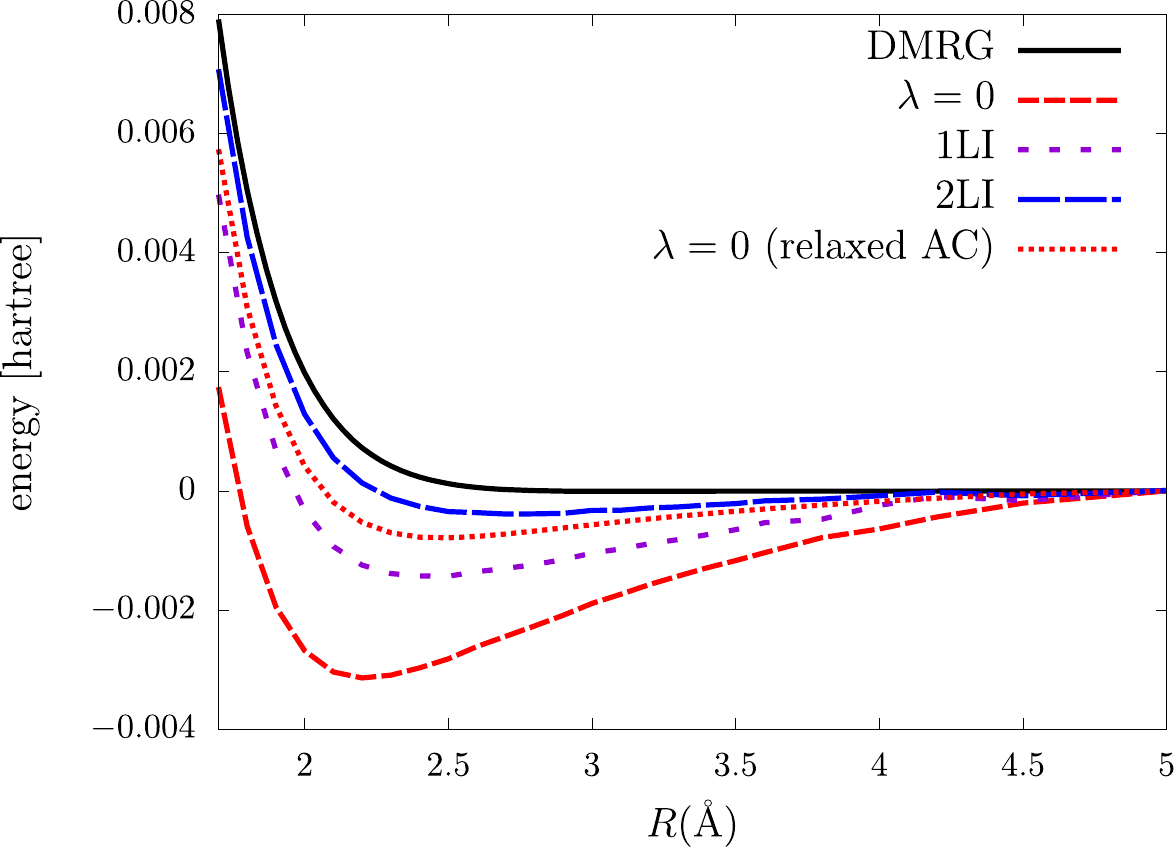}
}
\caption{Stretching energy $E(R)-E(R=5\;\mbox{\AA})$ curves of the Helium
dimer computed at both seniority-zero ($\lambda=0$) and higher-seniority
(through linear interpolations of the constrained AC integrand) levels of approximation. The relaxed seniority-zero ($\lambda=0$) curve is shown for analysis
purposes. The DMRG curve, which is bound (see
Table~\ref{tab:diss_He2}), is used as reference. See text for further details.}
\label{fig:diss_He2}
\end{figure}

This ``dilemma'' is also
reflected in the stretching energy curves of the Helium dimer (see
Fig.~\ref{fig:diss_He2} and Table~\ref{tab:diss_He2}). In comparison
with standard DMRG (which gives similar results to
FCI~\cite{zoboki2013linearized}), the reference ($\lambda=0$) seniority-zero calculation of the constrained
AC dramatically overbinds: The equilibrium bond distance is
underestimated by 0.9 {\AA} and the stretching energy is wrong by a
factor 10$^3$. This observation clearly shows the importance of the complementary
higher-seniority density-matrix correlation functional in this
context. Even though the 2LI approximation recovers a
substantial amount of the missing correlation energy at equilibrium (see
the bottom panel of Fig.~\ref{fig:Wc_allsystems} and
Table~\ref{tab:corr}), it fails in describing dynamical
correlations at longer bond distances, thus leading to an overestimation
by a factor 10$^2$ of the stretching energy. 
Better approximations to the higher-seniority correlation energy are definitely needed in this case.
For the sake of completeness, 
we verified numerically that, in the
dissociated $R=30$ {\AA} geometry, the AC
integrand equals zero for all $\lambda$ values (not shown), as expected for two
separate two-electron systems (each of them being described exactly by a
seniority-zero wave function). Note that the errors at $\lambda=0$ (seniority-zero limit) are
substantially reduced when switching from the constrained to the relaxed
AC, but they remain
significant. As expected from the comparison of Table~\ref{tab:diss_He2} (see the equilibrium energies) with Fig.~1 of Ref.~\cite{zoboki2013linearized},
where the dissociation curve of the Helium dimer has been computed (without
BSSE corrections) with a
variationally optimized
{\it antisymmetrized product of strongly orthogonal geminals} (APSG),
these errors would be further (and drastically) reduced by relaxing, in addition, the natural
orbitals constraint. An accurate description of Van der Waals
interactions would still require both
post-seniority-zero treatment and BSSE corrections~\cite{zoboki2013linearized}.
\\ 

In summary, using as reference a seniority-zero wave function that
reproduces the exact 1RDM (by analogy with DFT, where the Kohn--Sham
determinant is used 
for reproducing the
exact density) complicates the evaluation of the missing 
higher-seniority correlation energy, which is substantial, because the
corresponding correlation integrand does not vary linearly along the AC.

\begin{table}
\begin{center}
\begin{tabularx}{\columnwidth}{l*{1}{Y} l*{3}{Y}}
\hline He$_2$
 & \multicolumn{1}{c}{$R_{\rm eq}$ (\AA)}
  & \multicolumn{1}{c}{\vtop{\hbox{\strut Equilibrium}\hbox{\strut energy}}}
    & \multicolumn{1}{c}{\vtop{\hbox{\strut Stretching}\hbox{\strut energy}}}  \\
\hline
$\lambda = 0$ & 2.2 & -5.721112  & 0.003138 \\
$\lambda = 0$ (relaxed AC) & 2.5 & -5.736052 & 0.000786 \\
1LI & 2.5 & -5.761040 & 0.001440\\
2LI & 2.7 & -5.771703 & 0.000395\\
$\lambda = 1$ (DMRG) & 3.1 & -5.775196 & 0.000006 \\
\hline
\end{tabularx}
\caption{
Equilibrium bond distance, total energy at equilibrium, and stretching energy (in Hartree) of the
He$_2$ molecule computed in the cc-pVDZ basis at the constrained and
relaxed seniority-zero
($\lambda=0$) levels of
approximation (which is equivalent to perturbation theory through first
order) and beyond. In the latter (constrained) case, 1LI and 2LI 
energy corrections are applied [see
Eqs.~(\ref{eq:compfun_1LI_total_ener})
and (\ref{eq:compfun_2LI_total_ener})].
Comparison is made with DMRG. The stretching energy is evaluated as the difference between
the energies at $R = 5$~\AA\, and $R=R_{\rm eq}$. See text for further
details.
}
\label{tab:diss_He2}
\end{center}
\end{table}

\section{Conclusions and perspectives}\label{sec:conclusions}

An alternative exact formulation of RDMFT,
where the (one-electron reduced) density matrix is mapped onto an {\it ab initio} seniority-zero many-body wave
function, has been derived. The optimization of both the DOCI coefficients
and the natural orbitals has been discussed at the
formal level. Exact and approximate adiabatic connection (AC) formulas have been derived and
implemented in order to describe the complementary higher-seniority
density matrix functional correlation energy. Our proof-of-concept numerical calculations
clearly show that relying on a seniority-zero wave function that
reproduces the exact density matrix makes the description of
higher-seniority correlation energies nontrivial. The
latter are indeed substantial in this case, and the corresponding AC 
integrand has a pronounced curvature. As a result, neither second-order
perturbation theory nor a single-segment linear interpolation will be
adequate in this context.\\

In order to turn
the present approach into a practical computational method for modelling strongly correlated
molecular systems, various practical issues need to be addressed. First of all,
one should design density matrix functional approximations to the higher-seniority
interaction energy which is obtained by
subtracting the seniority-zero density matrix functional energy from
the regular full interaction one. The latter might be described with
standard functionals, such as the Piris natural orbital
functionals (PNOFs)~\cite{piris2014perspective,piris2014interacting}, possibly with perturbative
corrections~\cite{piris2017global}, in order to fully describe intrapair and
interpair correlations. 
Power-type functionals~\cite{sharma2008reduced} might also be used,
or functionals describing intergeminal correlation effects~\cite{mentel2012formulation,van2018non}.
Turning to the design of seniority-zero density matrix functional
approximations, using the
Richardson--Gaudin wave function as a model, as recently proposed by 
Johnson \etal~\cite{johnson2020richardsongaudin}, is an appealing idea. 
In order to evaluate numerically a seniority-zero interaction energy in the present context, the single particle energies would be determined
from the orbital occupation constraint while the pairing strength could
still be evaluated 
variationally, like in Ref.~\onlinecite{johnson2020richardsongaudin}.
Ultimately, we should be able to construct a complementary higher-seniority functional that depends explicitly on the natural
orbitals and their occupations, thus allowing for a self-consistent
computation of the natural orbitals and the seniority-zero wave
function [see Eqs.~(\ref{eq:final_sc_S0_eq}),
(\ref{eq:final_sc_S0_eq_hamil}), and (\ref{eq:final_sc_eq_nat_orb})].
The challenge, as one follows the above-mentioned strategies, is to ensure that
the two (fully and seniority-zero-only interacting) density matrix
functional approximations are merged in a consistent way. 
For example,
using PNOF5~\cite{piris2011natural} in this context would be irrelevant as it does not include
any higher-seniority correlation effect~\cite{pernal2013equivalence,piris2013interpair}.
However, the on-top density functionals used in Refs.~\onlinecite{hollett2016cumulant,elayan2022delta}
to recover the dynamical correlation missing in the $\Delta$NO approach could be used, as the on-top pair density is expressed as a functional of the natural orbitals and their occupations (see Eq.~34 of Ref.~\onlinecite{hollett2016cumulant}).
Note also that the resulting higher-seniority
functional should in principle vanish in the two-electron case.\\

Alternatively, 
a direct  
computation of the higher-seniority correlation energy, which then
becomes an implicit functional of the density matrix,
can be performed using
the AC formulas in Eqs.~(\ref{eq:final_AC_formula}) and
(\ref{eq:corr_int}).
In analogy
with recent works by Pernal~\cite{pernal2018electron} and then Vu~\etal~\cite{vu2019adiabatic}, the extended 
random phase approximation could be used for computing AC integrands. 
One could also follow a different AC path, 
where the seniority number gradually increases
until
the full configuration space is covered, 
in the spirit of 
Refs~\onlinecite{bytautas2011seniority,gebauer2013favorably}. 
Ideally, the
``potential'' (\ie, the single particle energies in the AC Hamiltonian)
should be adjusted along the AC so that the
natural orbital occupation constraint is fulfilled. 
Note that natural
orbitals and occupations could be extracted from a regular (and lower cost) NOFT 
calculation using PNOFs or power functionals, for example.
The above-mentioned implementation schemes should
obviously be explored further. This is left for future work.\\     

Finally, the computational burden of performing DOCI calculations
within our seniority-zero reformulation of RDMFT can be
bypassed
by considering recent advances in practical DOCI calculations,
such as variational 2RDM-based
approximations~\cite{vu2019adiabatic,head2020active,
alcoba2021variational,li2021challenges},
CIPSI~\cite{kossoski2021excited}, FCIQMC solvers~\cite{shepherd2016using},
a proper choice of Richardson-Gaudin states~\cite{fecteau2022near},
the pair parametric two-electron reduced density matrix approach~\cite{vu2020size},
or even using quantum computers~\cite{elfving2021simulating}. 
One can also truncate the configuration space by
considering a maximal degree of excitation~\cite{kossoski2022hierarchy}, or by defining an active space for each electron pair as in the $\Delta$NO method~\cite{elayan2022delta}.
pCCD and Ap1rog
are also appropriate candidates with mean-field cost~\cite{johnson2013size,limacher2013new,tecmer2014assessing,
stein2014seniority,henderson2015pair},
although they might break down if quadruple excitations are not negligible as they approximate the amplitudes $T_4$ by $T_2^2$.

\section{Acknowledgments}
The authors would like to thank Paul Ayers and Laurent Lemmens for
enlightening discussions. They also thank LabEx CSC (grant number:
ANR-10-LABX-0026-CSC), IdEx (University of Strasbourg, grant number:
W21RPD04), and ANR (grant number: ANR-19-CE07-0024-02 CoLab) for funding. 
S.Y. acknowledges support from the Interdisciplinary Thematic Institute
ITI-CSC via the IdEx Unistra (ANR-10-IDEX-0002).

\begin{appendices}
\appendix 
\numberwithin{equation}{section}
\setcounter{equation}{0}
\renewcommand{\thetable}{\Alph{section}}

\section{Hohenberg--Kohn theorem in the natural orbital space}\label{app:HKtheo}

The natural orbital basis is fixed in the following. Let us consider two
potentials $\{\varepsilon_p\}$ and $\{\varepsilon'_p\}$ that differ by
more than a constant, and the following {\it ground-state} seniority-zero interacting
Schr\"{o}dinger equations:
\be\label{eq:SchrEq_S0_one}
\left(\hat{W}^{\mathcal{S}_0}_{\rm
ee}+\sum_{p}\varepsilon_p\hat{n}_p\right)\ket{\Psi}=\mathcal{E}\ket{\Psi},
\ee  
and
\be\label{eq:SchrEq_S0_two}
\left(\hat{W}^{\mathcal{S}_0}_{\rm
ee}+\sum_{p}\varepsilon'_p\hat{n}_p\right)\ket{\Psi'}=\mathcal{E}'\ket{\Psi'}.
\ee  
If we assume that $\Psi=\Psi'$, then it comes from
Eqs.~(\ref{eq:SchrEq_S0_one}) and (\ref{eq:SchrEq_S0_two}) that
\be\label{eq:SchrEq_diff_pot}
\sum_{p}(\varepsilon'_p-\varepsilon_p)\hat{n}_p\ket{\Psi}=\left(\mathcal{E}'-\mathcal{E}\right)\ket{\Psi}.
\ee  
The ground-state wave function $\Psi$ can in principle be determined
from any
seniority-zero Slater determinant $\Phi_J$ that overlaps with $\Psi$ through a diffusion process:
\be
\ket{\Psi}\sim \lim_{\tau\rightarrow
+\infty}e^{-\hat{\mathcal{H}}\tau}\ket{\Phi_J},
\ee
where $\hat{\mathcal{H}}=\hat{W}^{\mathcal{S}_0}_{\rm
ee}+\sum_{p}\varepsilon_p\hat{n}_p$ is the
seniority-zero Hamiltonian in Eq.~(\ref{eq:SchrEq_S0_one}). Applying
$\hat{\mathcal{H}}$ repeatedly (through the exponential) will allow electrons to jump
from doubly occupied orbitals in $\Phi_I$ to unoccupied ones. Exchange-time-inversion
integrals [see Eq.~(\ref{eq:WS0})] are central in this process.
Thus we deduce the following decomposition:  
\be\label{eq:decomp_WF_S0_GS}
\ket{\Psi}=\sum_{I\in \mathcal{S}_0}C_I\ket{\Phi_I},
\ee
which is expected to cover the entire seniority-zero subspace. In other
words, all $C_I$ coefficients are expected to be non-zero, whatever the
index $I$ is. Therefore,
inserting Eq.~(\ref{eq:decomp_WF_S0_GS}) into
Eq.~(\ref{eq:SchrEq_diff_pot}) leads to 
\be\label{eq:diff_pot_nocc_constant}
\sum_{p}(\varepsilon'_p-\varepsilon_p)n^{\Phi_I}_{p}=\left(\mathcal{E}'-
\mathcal{E}\right), \forall I,
\ee
where $n^{\Phi_I}_{p}$ is the occupation of the natural orbital $p$ in
$\Phi_I$. If we now assign $N-2$ electrons to a fixed set of
natural orbitals
$\left\{p_i\right\}_{1\leq i\leq \frac{N}{2}-1}$ and let the  
remaining two electrons occupy any other orbital $q$, it comes from
Eq.~(\ref{eq:diff_pot_nocc_constant}):
\be\label{eq:HK_proof_constant_outside}
2(\varepsilon'_q-\varepsilon_q)&\overset{q\notin \left\{p_i\right\}}{=}&\left(\mathcal{E}'-
\mathcal{E}\right)-2\sum^{\frac{N}{2}-1}_{i=1}(\varepsilon'_{p_i}-\varepsilon_{p_i})
,
\ee
thus leading to
\be\label{eq:diff_pot_constant_outside}
\varepsilon'_q-\varepsilon_q&\overset{q\notin \left\{p_i\right\}}{=}&c, 
\ee 
where $c$ is a constant. As a result, applying
Eq.~(\ref{eq:diff_pot_nocc_constant}) to a different situation where none
of the $\left\{p_i\right\}$ orbitals are occupied leads to
\be\label{eq:Eprime_minus_E}
\mathcal{E}'-
\mathcal{E}=Nc.
\ee
If we now keep only two electrons within 
the $\left\{p_i\right\}$ orbital subspace, while redistributing the
remaining $N-2$ electrons among the other orbitals, we finally
obtain from Eqs.~(\ref{eq:diff_pot_nocc_constant}) and
(\ref{eq:diff_pot_constant_outside}):
\be
2(\varepsilon'_q-\varepsilon_q)&\overset{q\in \left\{p_i\right\}}{=}&
\left(\mathcal{E}'-
\mathcal{E}\right)-2\left(\frac{N}{2}-1\right)c,
\ee
or, equivalently [see Eq.~(\ref{eq:Eprime_minus_E})],
\be\label{eq:diff_pot_constant_inside}
\varepsilon'_q-\varepsilon_q\overset{q\in \left\{p_i\right\}}{=}c.
\ee
According to Eqs.~(\ref{eq:diff_pot_constant_outside}) and
(\ref{eq:diff_pot_constant_inside}), the two potentials differ by a
constant, which is impossible. Therefore $\Psi\neq \Psi'$.\\    

If we now assume, for simplicity, that seniority-zero ground-state energies
are not degenerate, then
\be
\expval{\hat{W}^{\mathcal{S}_0}_{\rm
ee}+\sum_{p}\varepsilon_p\hat{n}_p}_{\Psi}<
\expval{\hat{W}^{\mathcal{S}_0}_{\rm
ee}+\sum_{p}\varepsilon_p\hat{n}_p}_{\Psi'},
\ee 
and
\be
\expval{\hat{W}^{\mathcal{S}_0}_{\rm
ee}+\sum_{p}\varepsilon'_p\hat{n}_p}_{\Psi'}<
\expval{\hat{W}^{\mathcal{S}_0}_{\rm
ee}+\sum_{p}\varepsilon'_p\hat{n}_p}_{\Psi},
\ee 
according to the Rayleigh--Ritz variational principle. If we assume that
$\Psi$ and $\Psi'$ have the same natural orbital occupations, then
\be
0<\expval{\hat{W}^{\mathcal{S}_0}_{\rm
ee}}_{\Psi}-\expval{\hat{W}^{\mathcal{S}_0}_{\rm
ee}}_{\Psi'}<0,
\ee
which is impossible, thus establishing the one-to-one correspondence
between the ground-state natural orbital occupations
$\left\{n^\Psi_p\right\}$ and the potential $\{\varepsilon_p\}$.

\section{Stationarity condition for the natural orbitals}\label{app:proof}

The purpose of this appendix is to prove
Eq.~(\ref{eq:final_sc_eq_nat_orb}). We start from Eq.~(\ref{eq:orb_rot})
whose expansion through first order in the orbital rotation parameters
gives
\be
\varphi_{p({\bm
\kappa})}(\bfr)=
\varphi_p(\bfr)+\sum_{q}\kappa_{pq}\varphi_q(\bfr)+\mathcal{O}({\bm
\kappa}^2),
\ee
or, equivalently,
\be
\varphi_{r({\bm
\kappa})}(\bfr)=\varphi_r(\bfr)+\sum_{s<r}\kappa_{rs}\varphi_s(\bfr)-\sum_{s>r}\kappa_{sr}\varphi_s(\bfr)
+\mathcal{O}({\bm
\kappa}^2),
\nonumber\\
\ee
thus leading to
\be\label{eq:deriv_orb_kappa_pq}
\left.\dfrac{\partial \varphi_{r({\bm
\kappa})}(\bfr)}{\partial \kappa_{pq}}
\right|_{{\bm\kappa}=0}=\delta_{rp}\varphi_q(\bfr)-\delta_{rq}\varphi_p(\bfr).
\ee
As a result, the first term on the left-hand side of 
Eq.~(\ref{eq:station_cond_nat_orb}) can be written as follows [see
Eq.~(\ref{eq:kappa_dependent_1int})]: 
\be\label{eq_appendix:stat_cond_orb_term1}
\begin{split}
&\left.(\partial \bm{h}(\bm{\kappa})/\partial \kappa_{pq}\vert
{\bfn}^{\Psi^{\mathcal{S}_0}})
\right|_{{\bm\kappa}=0}
\\
&=2\sum_r
\left.\bra{\varphi_r}\hat{h}\ket{\partial \varphi_{r({\bm
\kappa})}/\partial \kappa_{pq}}
\right|_{{\bm\kappa}=0}
n^{\Psi^{\mathcal{S}_0}}_r
\\
&=2h_{pq}\left(n^{\Psi^{\mathcal{S}_0}}_p-n^{\Psi^{\mathcal{S}_0}}_q\right).
\end{split}
\ee
Turning to the second term on the left-hand side of
Eq.~(\ref{eq:station_cond_nat_orb}), we may rewrite the expectation
value for the seniority-zero interaction as [see Eq.~(\ref{kappa_dependent_2e_rep})]
\be
\begin{split}
&\bra{\Psi^{\mathcal{S}_0}}\hat{W}^{\mathcal{S}_0}_{\rm ee}(\bm{\kappa})
\ket{\Psi^{\mathcal{S}_0}}
=
\dfrac{1}{2}\sum_{p\neq q} \psh{pq}{pq}^{\bm\kappa} \expval{\hat{n}_p \hat{n}_q}_{\Psi^{\mathcal{S}_0}}
\\
&- \dfrac{1}{2}\sum_{p\neq q} \sum_\sigma \psh{pq}{qp}^{\bm\kappa}
\expval{\hat{n}_{p\sigma} \hat{n}_{q\sigma}}_{\Psi^{\mathcal{S}_0}}
\\
 &+ \sum_{pq} \psh{pp}{qq}^{\bm\kappa} \expval{\hat{P}_p^\dagger
\hat{P}_q}_{\Psi^{\mathcal{S}_0}},
\end{split}
\ee
or, equivalently [we use the simplified expression $\expval{\hat{n}_{p\sigma}
\hat{n}_{q{\sigma'}}}_{\Psi^{\mathcal{S}_0}}=({1}/{4})\expval{\hat{n}_p
\hat{n}_{q}}_{\Psi^{\mathcal{S}_0}}$ which holds for seniority-zero wave
functions],
\be\label{eq:simplified_S0_int_ener}
\begin{split}
&\bra{\Psi^{\mathcal{S}_0}}\hat{W}^{\mathcal{S}_0}_{\rm ee}(\bm{\kappa})
\ket{\Psi^{\mathcal{S}_0}}
\\
&=\dfrac{1}{4}\sum_{rs}
\left[2\psh{rs}{rs}^{\bm\kappa}-\psh{rs}{sr}^{\bm\kappa}\right](1-\delta_{rs}) \expval{\hat{n}_r \hat{n}_s}_{\Psi^{\mathcal{S}_0}}
\\
 &\quad+ \sum_{rs} \psh{rr}{ss}^{\bm\kappa} \expval{\hat{P}_r^\dagger
\hat{P}_s}_{\Psi^{\mathcal{S}_0}}
\\
&=\dfrac{1}{4}\sum_{rs}
\left[2\psh{rs}{rs}^{\bm\kappa}-\psh{rs}{sr}^{\bm\kappa}\right] \expval{\hat{n}_r \hat{n}_s}_{\Psi^{\mathcal{S}_0}}
\\
 &\quad+ \sum_{rs} \psh{rr}{ss}^{\bm\kappa}(1-\delta_{rs}) \expval{\hat{P}_r^\dagger
\hat{P}_s}_{\Psi^{\mathcal{S}_0}},
\end{split}
\ee
where we used the fact that $\expval{\hat{P}_r^\dagger
\hat{P}_r}_{\Psi^{\mathcal{S}_0}} = \dfrac{1}{4} \expval{\hat{n}_r
\hat{n}_r}_{\Psi^{\mathcal{S}_0}}$.
Since, according to Eq.~(\ref{eq:deriv_orb_kappa_pq}),
\be
\begin{split}
\dfrac{1}{2}\left.\partial \psh{rs}{
rs}^{\bm \kappa} 
/\partial \kappa_{pq}\right|_{{\bm\kappa}=0}
&=\delta_{rp}\psh{qs}{rs}-\delta_{rq}\psh{ps}{rs}
\\
&\quad+\delta_{sp}\psh{rq}{rs}-\delta_{sq}\psh{rp}{rs},
\end{split}
\ee

\be
\begin{split}
\dfrac{1}{2}\left.\partial \psh{rs}{
sr}^{\bm \kappa}
/\partial \kappa_{pq}\right|_{{\bm\kappa}=0}
&=
\delta_{rp}\psh{qs}{sr}-\delta_{rq}\psh{ps}{sr}
\\
&
\quad+\delta_{sp}\psh{rq}{sr}-\delta_{sq}\psh{rp}{sr},
\end{split}
\ee
and
\be
\begin{split}
\dfrac{1}{2}\left.\partial \psh{rr}{
ss}^{\bm \kappa}
/\partial \kappa_{pq}\right|_{{\bm\kappa}=0}
&=
\delta_{rp}\psh{qr}{ss}-\delta_{rq}\psh{pr}{ss}
\\
&
\quad+\delta_{sp}\psh{rr}{qs}-\delta_{sq}\psh{rr}{ps},
\end{split}
\ee
we obtain from Eq.~(\ref{eq:simplified_S0_int_ener}):
\be\label{eq_appendix:stat_cond_orb_term2}
\begin{split}
&
\left.\bra{\Psi^{\mathcal{S}_0}}\partial \hat{W}^{\mathcal{S}_0}_{\rm ee}(\bm{\kappa})/\partial \kappa_{pq}
\ket{\Psi^{\mathcal{S}_0}}
\right|_{{\bm\kappa}=0}
\\
&=\sum_r\left[2\psh{rp}{rq}-\psh{rp}{qr}\right]
\left[\expval{\hat{n}_r
\hat{n}_p}_{\Psi^{\mathcal{S}_0}}
-\expval{\hat{n}_r
\hat{n}_q}_{\Psi^{\mathcal{S}_0}}
\right]
\\
&\quad
+4\sum_r\psh{pq}{rr}
\\
&\qquad\times\left[
 (1-\delta_{rp})\expval{\hat{P}_r^\dagger
\hat{P}_p}_{\Psi^{\mathcal{S}_0}}
-(1-\delta_{rq})
\expval{\hat{P}_r^\dagger
\hat{P}_q}_{\Psi^{\mathcal{S}_0}}
\right].
\end{split}
\ee
Finally, by expressing the third term on the left-hand side of
Eq.~(\ref{eq:station_cond_nat_orb}) as follows: 
\be
\begin{split}
&
\left.\dfrac{\partial \overline{W}^{\mathcal{S}}({\bm \kappa},{\bfn
}^{\Psi^{\mathcal{S}_0}})}{\partial
\kappa_{pq}}
\right|
_{{\bm\kappa}=0}
=
\left.\dfrac{\partial
\overline{W}^{\mathcal{S}}\left(\left\{\varphi_{r({\bm
\kappa})}\right\},{\bfn
}^{\Psi^{\mathcal{S}_0}}\right)}{\partial
\kappa_{pq}}
\right|
_{{\bm\kappa}=0}
\\
&=
\sum_r
\int d\bfr
\left.\dfrac{\partial \varphi_{r({\bm
\kappa})}(\bfr)}{\partial \kappa_{pq}}
\right|_{{\bm\kappa}=0}
\dfrac{\delta \overline{W}^{\mathcal{S}}\left(\left\{\varphi_r\right\},{\bfn
}^{\Psi^{\mathcal{S}_0}}\right)}{\delta \varphi_r(\bfr)}
\\
&\equiv\sum_r 
\left\langle
\left.\dfrac{\partial \varphi_{r({\bm
\kappa})}}{\partial \kappa_{pq}}
\right|_{{\bm\kappa}=0}
\middle\vert
\dfrac{\delta \overline{W}^{\mathcal{S}}\left(\left\{\varphi_r\right\},{\bfn
}^{\Psi^{\mathcal{S}_0}}\right)}{\delta \varphi_r}
\right\rangle
\\
&=
\left\langle
\varphi_q
\middle\vert
\dfrac{\delta \overline{W}^{\mathcal{S}}\left(\left\{\varphi_p\right\},{\bfn
}^{\Psi^{\mathcal{S}_0}}\right)}{\delta \varphi_p}
\right\rangle
\\
&\quad-
\left\langle
\varphi_p
\middle\vert
\dfrac{\delta \overline{W}^{\mathcal{S}}\left(\left\{\varphi_q\right\},{\bfn
}^{\Psi^{\mathcal{S}_0}}\right)}{\delta \varphi_q}
\right\rangle
,
\end{split}
\ee
we recover Eq.~(\ref{eq:final_sc_eq_nat_orb}) from Eqs.~(\ref{eq_appendix:stat_cond_orb_term1}),
(\ref{eq_appendix:stat_cond_orb_term2}), (\ref{eq:station_cond_nat_orb})
and (\ref{eq:DM_fun_Fock_matrix}).

\section{PT2/PT3-based Pad\'{e} approximants to the AC integrand}\label{app:pade}

By analogy with the interaction-strength interpolation schemes that have been
developed in the context of DFT~\cite{seidl1999strictly,seidl1999strong,seidl2000simulation,vuckovic2016exchange,ernzerhof1996construction,mori-sanchez2006self}, we
propose two PT-based Pad\'{e} approximants to the AC
integrand. In the first one, which is referred to as PT2-Pad\'{e}
approximant and reads
\be
\begin{split}
\overline{\mathcal{W}}^{\mathcal{S},\lambda}
\overset{\mbox{PT2-Pad\'{e}}}{\approx}
&-\dfrac{\mathcal{W}_1\mathcal{W}^{\lambda=1}}{\mathcal{W}^{\lambda=1}-\mathcal{W}_1}
\\
&\times\left(1-\dfrac{1}{1-\dfrac{\left(\mathcal{W}^{\lambda=1}-\mathcal{W}_1\right)}{\mathcal{W}^{\lambda=1}}\lambda}\right)
,
\end{split}
\ee
where we denote
$\mathcal{W}^{\lambda=1}=\overline{\mathcal{W}}^{\mathcal{S},\lambda=1}$
and $\mathcal{W}_1=\left.\partial
\overline{\mathcal{W}}^{\mathcal{S},\lambda}/\partial
\lambda\right|_{\lambda=0}$,
the $\lambda\rightarrow0$ and $\lambda\rightarrow1$ limits
of the AC integrand are described exactly 
through first order (in $\lambda$) and zeroth order (in $\lambda-1$),
respectively. At this level of approximation, the energy (which is
obtained after integration over $\lambda$) is exact through second order in $\lambda$,
hence the name of the approximant.\\

The second approximant, which is referred to as PT3-Pad\'{e} and relies exclusively on the
$\lambda\rightarrow 0$ limit of the AC, reads
\be
\overline{\mathcal{W}}^{\mathcal{S},\lambda}\overset{\mbox{PT3-Pad\'{e}}}{\approx}-\dfrac{\mathcal{W}^2_1}{\mathcal{W}_2}\left(1-\dfrac{1}{1-\dfrac{\mathcal{W}_2}{\mathcal{W}_1}\lambda}\right),
\ee
where $\mathcal{W}_2=\frac{1}{2}\left.\partial^2
\overline{\mathcal{W}}^{\mathcal{S},\lambda}/\partial
\lambda^2\right|_{\lambda=0}$. It reproduces the correct expansion of
the integrand through second order in
$\lambda$,
\be
\overline{\mathcal{W}}^{\mathcal{S},\lambda}\overset{\rm{PT3}}{\approx}\lambda
\mathcal{W}_1+\lambda^2\mathcal{W}_2,
\ee
and therefore is exact through third order in
$\lambda$ for the energy (after integration over $\lambda$), 
\be
E\overset{\rm PT3}{\approx}\langle \hat{H}
\rangle_{\Psi^{\mathcal{S}_0}}+\dfrac{1}{2}\mathcal{W}_1+\dfrac{1}{3}\mathcal{W}_2,
\ee
hence the
name of the approximant. 
For completeness, we derived the PT2-Pad\'{e} and PT3-Pad\'{e} energies, that read
\begin{eqnarray}
E &\overset{\text{PT2-Pad\'{e}}}{\approx}&\langle \hat{H}
\rangle_{\Psi^{\mathcal{S}_0}}
-  \dfrac{\mathcal{W}_1 \mathcal{W}^{\lambda=1}}{\mathcal{W}^{\lambda=1} - \mathcal{W}_1}
\nonumber \\
&&- \dfrac{\mathcal{W}_1 (\mathcal{W}^{\lambda=1})^2}{(\mathcal{W}^{\lambda=1} - \mathcal{W}_1)^2}
\ln\left| 1 - \dfrac{\mathcal{W}^{\lambda=1} - \mathcal{W}_1}{\mathcal{W}^{\lambda=1}}\right| \nonumber \\
\end{eqnarray}
and
\begin{eqnarray}
E \overset{\text{PT3-Pad\'{e}}}{\approx}\langle \hat{H}
\rangle_{\Psi^{\mathcal{S}_0}}
- \dfrac{\mathcal{W}_1^2}{\mathcal{W}_2} - \dfrac{\mathcal{W}_1^3}{\mathcal{W}_2^2} \ln \left| 1 - \dfrac{\mathcal{W}_2}{\mathcal{W}_1}\right|.
\end{eqnarray}
In the present work, $\mathcal{W}_1$ and 
$\mathcal{W}_2$ have been evaluated numerically,
for analysis purposes. In practice, exact analytical expressions might
be used instead [see Eq.~(\ref{eq:deriv_AC}) for $\mathcal{W}_1$]. Regarding $\mathcal{W}_2$,
according to
Eqs.~(\ref{eq:Hamilt_along_AC}) and (\ref{eq:first_deriv_integrand_derivHamil}), we have 
\be\label{eq:full_exp_second_deriv_integrand}
\begin{split}
\dfrac{1}{2}\dfrac{\partial^2 \overline{\mathcal{W}}^{\mathcal{S},\lambda}}{\partial
\lambda^2}&=
\expval{\hat{\mathcal{V}}^\lambda}_{\frac{\partial
\Psi_0^\lambda}{\partial\lambda}}
+
\mel{
\Psi_0^\lambda}{
\hat{\mathcal{V}}^\lambda
}
{\frac{\partial^2 
\Psi_0^\lambda}{ \partial \lambda^2}} 
\\
&\quad+
\mel{
\Psi_0^\lambda
}{\dfrac{\partial^2
\hat{\varepsilon}^\lambda}{\partial \lambda^2}}
{\frac{\partial 
\Psi_0^\lambda}{ \partial \lambda}}, 
\end{split}
\ee
where we used the shorthand notations 
$\hat{\mathcal{V}}^\lambda=\hat{\mathcal{V}}+\frac{\partial\hat{\varepsilon}^\lambda}{\partial\lambda}$
and $\hat{\varepsilon}^\lambda=\sum_p
\varepsilon_p^\lambda \hat{n}_p$.
Since the last term on the right-hand side of
Eq.~(\ref{eq:full_exp_second_deriv_integrand}) vanishes, according to
the natural orbital occupations constraint,
\be\label{eq:PT3_ener_second_order_pert_cancel_out}
2\mel{
\Psi_0^\lambda
}{\dfrac{\partial^2
\hat{\varepsilon}^\lambda}{\partial \lambda^2}}
{\frac{\partial 
\Psi_0^\lambda}{ \partial \lambda}}= 
\sum_p\dfrac{\partial^2
{\varepsilon}_p^\lambda}{\partial \lambda^2}\dfrac{\partial
\expval{\hat{n}_p}_{\Psi_0^\lambda}}{\partial\lambda}
=0,
\ee 
the second-order derivative of the constrained AC integrand can be
written more explicitly in perturbation theory as follows~\cite{srDFT_densitymatrixformulation}, 
\be\label{eq:full_2nd_deriv_integrand_exp}
\begin{split}
&\dfrac{1}{6}\dfrac{\partial^2 \overline{\mathcal{W}}^{\mathcal{S},\lambda}}{\partial
\lambda^2}=
\\
&\sum_{I>0} \sum_{J>0}
\dfrac{\bra{\Psi_0^\lambda} \hat{\mathcal{V}}^\lambda \ket{\Psi_J^\lambda}
\bra{\Psi_J^\lambda} \hat{\mathcal{V}}^\lambda \ket{\Psi_I^\lambda}
\bra{\Psi_I^\lambda} \hat{\mathcal{V}}^\lambda \ket{\Psi_0^\lambda}}{(\mathcal{E}^\lambda_0 - \mathcal{E}^\lambda_I)(\mathcal{E}^\lambda_0 - \mathcal{E}^\lambda_J)} \\
& -\bra{\Psi_0^\lambda} \hat{\mathcal{V}}^\lambda\ket{\Psi_0^\lambda}
\sum_{I>0} \dfrac{| \bra{\Psi_0^\lambda} \hat{\mathcal{V}}^\lambda \ket{\Psi_I^\lambda} |^2}{(\mathcal{E}^\lambda_0 - \mathcal{E}^\lambda_I)^2}
\\
&
+\dfrac{1}{3}\sum_{I>0}\dfrac{\bra{\Psi_0^\lambda}
\hat{\mathcal{V}}^\lambda \ket{\Psi_I^\lambda}\bra{\Psi_I^\lambda}\left(\frac{\partial^2
\hat{\varepsilon}^\lambda}{\partial \lambda^2}\right)
\ket{\Psi_0^\lambda}}{\mathcal{E}^\lambda_0 - \mathcal{E}^\lambda_I}.
\end{split}
\ee
Note that the last term on the right-hand side of Eq.~(\ref{eq:full_2nd_deriv_integrand_exp})
originates from the fact that, unlike in regular perturbation theory, the perturbation
operator contains second- (and higher-) order
contributions~\cite{srDFT_densitymatrixformulation,angyan2008rayleigh,fromager2011analysis}:  
\be
\begin{split}
\left(\lambda+\delta\right)\hat{\mathcal{V}}+\hat{\varepsilon}^{\lambda+\delta}
&=
\lambda\hat{\mathcal{V}}+\hat{\varepsilon}^\lambda+
\delta\hat{\mathcal{V}}^\lambda
\\
&
\quad+\dfrac{\delta^2}{2}\dfrac{\partial^2
\hat{\varepsilon}^\lambda}{\partial
\lambda^2}+\mathcal{O}\left(\delta^3\right).
\end{split}
\ee
When rewritten as follows in terms of the first-order wave function, it actually
cancels out, as readily seen from Eq.~(\ref{eq:PT3_ener_second_order_pert_cancel_out}):
\be
&\dfrac{1}{3}\mel{\dfrac{\partial
\Psi_0^\lambda}{\partial\lambda}}{\left(\dfrac{\partial^2
\hat{\varepsilon}^\lambda}{\partial
\lambda^2}\right)}{\Psi_0^\lambda}=0.
\ee
As a result, $\mathcal{W}_2$ can be evaluated exactly solely from the
first-order wave function, like in regular PT3:
\be
\dfrac{1}{6}\dfrac{\partial^2 \overline{\mathcal{W}}^{\mathcal{S},\lambda}}{\partial
\lambda^2}&=
\expval{\hat{\mathcal{V}}^\lambda}_{\frac{\partial
\Psi_0^\lambda}{\partial\lambda}}
-\expval{\hat{\mathcal{V}}^\lambda}_{\Psi_0^\lambda}
\left\langle{\frac{\partial
\Psi_0^\lambda}{\partial\lambda}}
\middle\vert
{\frac{\partial
\Psi_0^\lambda}{\partial\lambda}}\right\rangle.
\ee


\section{Convergence of the DMRG energy}\label{app:convergence_DMRG}

The DMRG sweep energy for a given number of renormalized states $m$ has a convergence criterion
of $10^{-10}$ Eh.
To assess the convergence with respect to $m$, we did several calculations for different values of $m$
reported in Table~\ref{tab:DMRG}.
\begin{table}\label{tab:DMRG}
\caption{DMRG energy in Hartree with respect to the number of renormalized states $m$.}
\begin{tabular}{|l|l||l|l|}
  \hline
H$_8$ & ($R = 1.8$ \AA)  &   He$_2$  & ($R = 3.1$ \AA) \\
  $m$ & DMRG energy & $m$ & DMRG energy\\
  \hline
100& -4.125716&20 &-5.775195\\
200& -4.129171&40 &-5.775195\\
300& -4.129746&100&-5.775195\\
400& -4.129900&200&-5.775195\\
500& -4.129949&300&-5.775195\\
600& -4.129969&400&-5.775195\\
700& -4.129978&&\\
800& -4.129982&&\\
1200&-4.129985&&\\
1600&-4.129986&&\\
  \hline
\end{tabular}
\end{table}
According to Table~\ref{tab:DMRG}, $m = 400$ is more than enough for the Helium dimer, for which even 20 states are sufficient.
For H$_8$ in the cc-pVDZ basis, the DMRG energy is converged up to $10^{-5}$ Hartree at $m=800$,
which we considered large enough to neglect the effect of orbital ordering.

\section{First- and second-order derivatives of the correlation integrand}\label{app:derivatives}

The first-order derivative of the correlation integrand
at $\lambda = 0$ is required
to compute the energy in the PT3 and PT2-Pad\'e approximations,
in addition to the second-order derivative at $\lambda = 0$ for the PT3-Pad\'e approximations.
They have been estimated by fitting the correlation integrand using the implementation of the nonlinear least-squares Marquardt-Levenberg algorithm in
gnuplot, in between $\lambda = 0$ and $\lambda = 0.2$.
A polynomial of degree 2, i.e. $a \lambda^2 + b\lambda$, has been considered
such that
\begin{eqnarray}
\left.
\dfrac{\partial \overline{\mathcal{W}}^{\mathcal{S},\lambda}}{\partial \lambda}\right|_{\lambda = 0}
&=& b,
\end{eqnarray}
and
\begin{eqnarray}
\left.
\dfrac{\partial ^2\overline{\mathcal{W}}^{\mathcal{S},\lambda}}{\partial \lambda^2}\right|_{\lambda = 0}
&=& a.
\end{eqnarray}
The values of $a$ and $b$ are reported in Table~\ref{tab:fit}.
\begin{table}\label{tab:fit}
\caption{Parameters (in Hartree) obtained by fitting the correlation integrand with the polynomial $a\lambda^2 + b\lambda$ in between $\lambda = 0$ and $\lambda = 0.2$.}
\begin{tabular}{l|l|l}
  \hline
System & $a$ &
  $b$\\
  \hline
  H$_4$ ($R = 0.9$ \AA) & 0.0572817 & -0.112019 \\
  H$_4$ ($R = 3.4$ \AA) & 0.74256 & -1.11383 \\
  He$_2$ ($R=3.1$ \AA) & 0.0997618 & -0.185009 \\
  H$_8$ ($R = 1.8$ \AA) & 1.7061 & -1.97377 \\
  \hline
\end{tabular}
\end{table}

\end{appendices}

\clearpage


\newcommand{\Aa}[0]{Aa}

\end{document}